\RequirePackage{fix-cm}
\documentclass[twocolumn]{svjour3}          
\smartqed  
\usepackage{comment}
\usepackage{graphicx}
\usepackage{amsmath, amssymb, amsfonts}
\usepackage{algorithm}
\usepackage{algorithmic}
\usepackage{listings}
\usepackage{url}
\usepackage{tcolorbox}
\usepackage{cite}
\usepackage{xcolor, soul}
\sethlcolor{green}
\usepackage{ulem}
\PassOptionsToPackage{hyphens}{url}
\usepackage{hyperref}
\usepackage[flushleft]{threeparttable}
\usepackage{listings}
\usepackage{tcolorbox}
\tcbuselibrary{skins}
\usetikzlibrary{shadings}
\usepackage{float}
\usepackage{pifont}
\usepackage{hyphenat}
\usepackage{makecell} 
\usepackage{colortbl}
\usepackage{float}
\newfloat{listing}{htbp}{lol} 

\definecolor{headercolor}{rgb}{0,0,0} 

\newtheorem{mydef}{Definition}
\begin{document}

\title{Finding Software Vulnerabilities in Open-Source C Projects via Bounded Model Checking}


\author{Janislley Oliveira de Sousa  \and
        Bruno Carvalho de Farias \and
        Thales Araujo da Silva \and
        Eddie Batista de Lima Filho  \and
        Lucas C. Cordeiro
}


\institute{Janislley Sousa \at
              Federal University of Amazonas and SIDIA Institute of Science and Technology, Manaus, Brazil \\
              \email{janislley.sousa@sidia.com}           
           \and
           Bruno Farias \at
              University of Manchester, Manchester, UK \\
              \email{bruno.farias@manchester.ac.uk} 
           \and
           Thales Silva \at
              Federal University of Amazonas, Manaus, Brazil \\
              \email{thales.tas@gmail.com} 
           \and
           Eddie Filho \at
              Federal University of Amazonas, Brazil and TPV Technology, Manaus, Brazil \\
              \email{eddie.filho@tpv-tech.com} 
           \and
           Lucas Cordeiro \at
               University of Manchester, Manchester, UK and Federal University of Amazonas, Brazil \\
              \email{lucas.cordeiro@manchester.ac.uk} 
}

\date{Received: date / Accepted: date}

\maketitle

\begin{abstract}
Computer-based systems have solved several domain problems, including industrial, military, education, and wearable. Nevertheless, such arrangements need high-quality software to guarantee security and safety as both are mandatory for modern software products. We advocate that bounded model-checking techniques can efficiently detect vulnerabilities in general software systems. However, such an approach struggles to scale up and verify extensive code bases. Consequently, we have developed and evaluated a methodology to verify large software systems using a state-of-the-art bounded model checker. In particular, we pre-process input source-code files and guide the respective model checker to explore them systematically. Moreover, the proposed scheme includes a function-wise prioritization strategy, which readily provides results for code entities according to a scale of importance. Experimental results using a real implementation of the proposed methodology show that it can efficiently verify large software systems. Besides, it presented low peak memory allocation when executed. We have evaluated our approach by verifying twelve popular open-source C projects, where we have found real software vulnerabilities that their developers confirmed.
\end{abstract}

\section{Introduction}
\label{intro}

In our society, problems of several domains (e.g., industrial, military, and education) have been solved by applying computer-based systems, which usually need high-quality software to satisfy a set of safety and security aspects~\cite{CordeiroFB20}. Furthermore, critical embedded systems, such as those in the industrial domain, impose several \textit{robustness} and \textit{security} properties~\cite{BarretoCF11, CordeiroF16}. According to a system's requirements, such properties must be verified and validated at the early development stages; otherwise, failures may lead to catastrophic situations, e.g., loss of financial resources and lives~\cite{CordeiroFB20}. As an outcome, software validation and verification techniques are essential for developing robust systems with high dependability and reliability requirements, ensuring that user requirements and a specific behavior are met~\cite{AlshmranyBBCKMM22}.

In critical domains, software must be carefully tested so that violations do not take place during execution in real scenarios~\cite{Myers2004}. For instance, in the C programming language~\cite{Kernighan2006}, widely used to develop critical software, e.g., operating systems and drivers, execution of unsafe code might lead to undefined behaviors~\cite{related-2}. This is a common cause of memory problems, including buffer overflows and double-free violations~\cite{cordeiro2011smt}. Furthermore, these errors are some of the main threats to software security~\cite{Veen2012} since attackers can exploit them to execute malicious code. Such an aspect is even worse in open-source software as the same attackers can quickly read code and easily find vulnerable spots. In addition, as this kind of project tends to be widely used by the general public, one can even question its very nature. Therefore, developers must use the available resources to validate source code more often. 

Also, new advanced tools should be further developed to improve software security~\cite{hoepman2007increased,ohm2020backstabber}. For instance, to demonstrate the importance of software validation and verification, one could mention the case that involved Log4j~\cite{log4j}. In short, data was printed out or logged into a file that might be used to take over a server. It could be done because Log4j permitted the injection of external logging text, whose format and content could be chosen through look-ups. This way, sensitive data could be leaked, or a network connection could be made to acquire and run malicious code.

Several schemes have been proposed to detect vulnerabilities in C programs. For instance, fuzzing~\cite{Bohme2017,Godefroid20,AldughaimAGFC23} techniques, including black-, grey- or white-box, have been widely used to exploit randomly-created program inputs that lead to unexpected behaviors. Static analysis tools that check C programs~\cite{CadarDE08,related-1, Clarke2004, Gadelha2019}, concerning safety properties, are another example of approaches that aim to find violations in software: widely known instances are CPPCheck and Flawfinder~\cite{PereiraCF20}. Code sanitizers such as Google's Address Sanitizer are other important tools for detecting issues in C programs~\cite{SerebryanyBPV12}. Semi-formal approaches combining static and dynamic verification have also been proposed~\cite{CordeiroFCM09}. However, such techniques struggle to scale up and verify large software systems commonly found in open-source projects. 

Recently, Cook {\it et al.}~\cite{Cook2020} showed the use of bounded model checkers to triage the severity of security bugs at the cloud service provider from Amazon Web Services (AWS). However, this approach's several abstractions might make the underlying bounded model checker miss traces and not automatically falsify spurious ones. Bounded model checking (BMC), making use of Boolean Satisfiability (SAT) or Satisfiability Modulo Theories (SMT), is an interesting approach that proved to be one of the most feasible options for software verification, being able to tackle a myriad of different systems as long as suitable models for them are available~\cite{CordeiroFB20}. In this respect, the efficient SMT-based bounded model checker (ESBMC), a practical example that employs BMC, has been applied to verify general software, which also includes digital filters~\cite{AbreuGCFS16}, controllers~\cite{BessaICF14,ChavesIBCF19}, and unmanned aerial vehicles~\cite{ChavesBIFCF18}. These studies create an abstract model of an underlying digital system, conservatively approximating its semantics and defining safety properties related to a given behavior. However, despite all the advantages of BMC, which make it a likely option for practical applications, it also struggles to handle large code bases.

One may also notice that large software systems are frequently composed of many elements declared in several source files, usually split across different directories. This brings another problem when static analysis tools are applied to this system: software model checkers usually verify a single file using a predetermined entry point~\cite{beyer2022progress}. The LSVerifier tool was recently developed to leverage the bounded model checking technique for effectively detecting security vulnerabilities in open-source C  software. This tool introduced an innovative approach to verifying open-source software and identifying issues that could lead to software vulnerabilities~\cite{eea4463fa55f40ba8695d7a600f5dbdd}. However, to handle large pieces of software with many files, a scenario typically found in open-source applications, it is necessary to verify each one at once and then change the current entry point when required. In addition, there can be elements with different priorities that must be handled appropriately. 

The last paragraphs inspired this study, i.e., proposing a pragmatic approach to verify large software systems: state-of-the-art bounded model checkers and parameters to be configured by users according to the vulnerability classes they want to check. We systematically guide an underlying verifier through source-code files to recursively explore threats in entire source-code directories or specific locations, e.g., functions, according to a pre-defined priority. As reported in our experimental evaluation, this allowed us to reduce computational resources and find bugs in large software projects, which the respective maintainers later confirmed. Furthermore, once a counterexample is identified, a user-friendly report is generated, showing information about the respective violation, such as file and target function, code line where it occurred, and the identified vulnerability. 

Our original contribution is a novel methodology that combines input-code analysis and BMC techniques to detect and evaluate software vulnerabilities in large software systems, a condition usually found in open-source C software projects. A real implementation of it was used during the mentioned experiments. Other significant contributions of our work are as follows:

\begin{itemize}
\item  A robust verification tool combining input-code analysis, prioritized function analysis, and BMC techniques to detect software vulnerabilities in large software systems, LSVerifier v0.3.0, making it well-suited for application in open-source development projects;
\item An evaluation structure that assesses findings in a detailed and user-friendly report, which can be used for problem reproduction and correction;
\item An in-depth evaluation of our approach over a large set of open-source applications, where experimental results show that it can find real software vulnerabilities, e.g., overflows, array-out-of-bounds, division by zero, and pointer-safety violations.
\item A literature review of the existing techniques applied to find software vulnerabilities in open-source applications, which shows gaps regarding methods and approaches and establishes research premises.
\end{itemize} 

It is worth noticing that software developers have also confirmed many vulnerabilities reported by LSVerifier. Moreover, even the ones not formally recognized revealed doubtful practices by developers, which could weaken the current open-source development process. Besides, although LSVerifier's design is somewhat new, its implementation is mature enough to handle large and complex open-source software such as Wireshark, VLC, and Cmake, thus paving the way to whole-system exploration approaches. The obtained experimental results show that our vulnerability-analysis method is feasible and can be helpful to the open-source software community. It also proved to be an important tool for checking the security of third-party libraries.

This paper is organized as follows. Section~\mbox{\ref{sec:background}} provides an overview of the background knowledge in vulnerability verification for large software systems, focusing on open-source software using BMC techniques. Section~\mbox{\ref{sec:method}}, in turn, presents our proposed method to verify large software systems using BMC. Then, Section \ref{sec:imple} describes a real implementation that can verify software in a completely automated manner, also disclosing all the necessary resources and specific algorithms. Following that, Section~\mbox{\ref{sec:experimental}} presents the experimental results obtained during this study, comparing its applicability for open-source projects.  Section~\mbox{\ref{sec:related-work}} discusses the studies published within the last ten years to verify software vulnerabilities in large C/C++ programs. Finally, Section~\mbox{\ref{sec:conclusion}} concludes this work and discusses its contributions and future studies.

\section{Background}
\label{sec:background}

As a formal definition, a software vulnerability is a security flaw, glitch, or weakness found in code that could be exploited by an attacker, leading, for instance, to sensitive-data leak or execution of malicious instructions~\cite{nistir8011}. In addition, the common weakness enumeration (CWE) community~\cite{cweWebsite} identifies the most common vulnerabilities associated with the C/C++ programming language~\cite{cwe}. Here, we describe ten vulnerability categories that we consider in this study.

\begin{mydef}
	\label{BufferOverflow}
	(\textit{Buffer Overflow}) This vulnerability is characterized by copying data from one buffer to another without checking whether the former fits within the latter, independently of memory location (i.e., heap, stack, etc.). Consequently, data in adjacent memory addresses becomes corrupted, which attackers can exploit to trigger crash events, incorrect program behavior, information leakage, or the execution of malicious code~\cite{cwe}. It is categorized under CWE-120 and can be defined as an unchecked buffer copy operation where the input size is not properly validated.
\end{mydef}

\begin{mydef}
	\label{ArithmeticOverflow}
	(\textit{Arithmetic Overflow}) This vulnerability is defined by an arithmetic operation's result surpassing the maximum capacity of its assigned data type. It is triggered when computations yield an integer overflow or wraparound, contradicting the assumption that the resultant value will invariably exceed the original. Such miscalculations can expose additional weaknesses, especially when used for resource allocation or execution control~\cite{cwe}. It is categorized under CWE-190 and defined as integer overflow or wraparound.
\end{mydef}

\begin{mydef}
	\label{InvalidPointer}
	(\textit{Invalid Pointer}) This vulnerability category includes the dereferencing of uninitialized (or null) pointers and the deallocation of memory using uninitialized or invalid pointers. When dereferencing an uninitialized or null pointer, different issues can occur, including crashes, data corruption, and unexpected program behavior. This is closely related to CWE-476. Deallocating memory through uninitialized or invalid pointers can result in memory corruption and program instability. This issue is commonly related to CWE-416, which primarily addresses use-after-free scenarios. Both forms of improper pointer usage can lead to unpredictable program behavior and may be exploited by attackers~\cite{cwe}.
\end{mydef}

\begin{mydef}
	\label{ImproperBufferAccess}
	(\textit{Improper Buffer Access}) This vulnerability occurs when software employs a sequential operation to read or write a buffer with an incorrect length value. As a result, memory outside of a buffer's boundaries can be accessed. This way, an attacker can gain access to sensitive data or even execute arbitrary code~\cite{cwe}. This software vulnerability falls under CWE-119 and is defined as an improper restriction of operations within the bounds of a memory buffer.
\end{mydef}

\begin{mydef}
    \label{NullPointerDerefence}
    (\textit{Null Pointer Dereference}) This vulnerability occurs when an application dereferences a null pointer, often due to race conditions or programming errors. Typically, this results in a program crash or an abrupt exit. An attack using it can aim at service denial~\cite{cwe}. It is categorized under CWE-476.
\end{mydef} 

\begin{mydef} 
    \label{DoubleFree} (\textit{Double Free}) This vulnerability occurs when a program calls the \textit{free()} function on the same pointer more than once, potentially leading to memory corruption and crashes. As a result, an attacker could gain unauthorized access to this memory buffer and execute arbitrary code or trigger program crashes~\cite{cwe}. This vulnerability is categorized under CWE-415.
\end{mydef} 

\begin{mydef} 
    \label{Dividebyzero} 
    (\textit{Division by zero}) This vulnerability occurs when a program attempts to divide a number by zero. If not handled correctly, it can cause program crashes or unexpected behavior. Moreover, in some cases, attackers may exploit it to trigger security issues~\cite{cwe}. It is categorized under CWE-369.
\end{mydef} 

\begin{mydef} 
    \label{outofbounds} 
    (\textit{Array bounds violated}) This vulnerability occurs when a program attempts to access an array element at an invalid index, either below zero or beyond an array's length, leading to data corruption, crashes, or unauthorized code execution~\cite{cwe}. It is categorized under CWE-787 and defined as modifying an index or performing pointer arithmetic that accesses a memory location outside a buffer's boundaries.
\end{mydef} 

\begin{mydef} 
    \label{sameobjectvio} 
    (\textit{Pointer arithmetic violation}) This vulnerability occurs when a product employs pointer arithmetic (e.g., subtraction, comparison) to ascertain size. It can lead to the same object violation when pointers from different memory blocks are used. In C programming, pointer arithmetic is commonly used to navigate arrays or compare memory positions, assuming the pointers involved reference the same allocated memory block. Issues arise when arithmetic operation is attempted between pointers not pointing to the same memory block, leading to undefined and unreliable results. It is categorized under CWE-469 and defined as using pointer subtraction to determine size.

\begin{mydef}
	\label{AssertionFailure}
	(\textit{Assertion violation}) This vulnerability occurs when a condition provided to the function \textit{assert} is not satisfied during program execution. A reachable assertion failure suggests the existence of a program execution path that leads to the location of the assertion, where the value of variables does not meet the expected conditions. Assertions are often used to verify that program variables remain within user-defined bounds. An assertion failure may reveal logical errors that could be exploited, resulting in unpredictable behavior or system crashes. This type of vulnerability is associated with CWE-617, where assertions are expected to hold during normal execution~\cite{cwe}.
\end{mydef}
\end{mydef} 

\subsection{Bounded Model Checking}
\label{sec:esbmc}

BMC techniques, based on SAT~\cite{Biere2009} or SMT~\cite{Moura2009}, have been successfully employed to verify single- and multi-threaded code, aiming to find subtle bugs in real software~\cite{Clarke2004, Merz2012, Gadelha2019, gadelha2021esbmc, Byer2015, Carter2016}. The general idea behind BMC is to check the negation of a given property at a specific depth. Formally, given a transition system $M$, a property $\phi$, and a limit of iterations $k$, BMC unfolds a target program $k$ times and converts it into a verification condition $\psi$, such that the latter is \textit{satisfiable} \textit{iff} $\phi$ has a counterexample of depth less than or equal to $k$.

BMC techniques can falsify properties up to a given depth $k$. However, they cannot prove system correctness unless the value of $k$ that unwinds all loops and recursive functions to their maximum possible depths is known. Consequently, BMC restricts the regions of data structures that can be visited and the number of related loop iterations. 

BMC limits the state space to be explored during verification so that fundamental application errors can be found~\cite{Clarke2004, Merz2012, Gadelha2019, gadelha2021esbmc, Ivancic2005}. Nevertheless, BMC tools are susceptible to exhaustion of time or memory limits when verifying programs with loop bounds that are too large. One example of a BMC tool is ESBMC \cite{Gadelha2019}, which can falsify predefined and user-defined safety properties \cite{cordeiro2011smt}.

\subsection{Software Vulnerabilities in Large Software Systems: Open-Source Projects}
\label{sec:open-source}

 As awareness regarding software risks increases, various vulnerability analysis tools are developed~\cite{seacord_2014}. At the same time, considerable effort is often made to research new techniques and approaches specifically for large software systems, especially open-source software. It happens because the latter is highly susceptible to security threats~\mbox{\cite{wen2017software, plate2015impact, muegge2018time}} due to its collaborative and more relaxed development process. 
 
 Besides, the number of new software vulnerabilities discovered in the last ten years increased $270$\%: more than $15,000$ new occurrences were reported from $2010$ to $2021$, according to the National Institute of Standards and Technology \mbox{\cite{nist2021}}. In addition, many were found in projects that used open-source code, making their security verification critical.

According to Xiao {\it et al.} \cite{xiao2014social}, security vulnerabilities are a significant challenge when creating applications with open-source code. Their research investigated several social factors that impact developers' adoption decisions based on a multidisciplinary field of research called \textit{diffusion of innovations}. Its results show that security tools can force developers to build more secure software systems by helping them detect or fix vulnerabilities in source code during implementation phases. In addition, conditions such as concerning behavior and lack of understanding regarding the consequences of security failures were identified in those whose primary activity is code writing. Moreover, while most open-source software projects have large communities contributing to their growth, some are not regularly maintained, which favors security faults.

Jing Zou {\it et al.}~\mbox{\cite{zou2019research}} pointed out a critical perspective: open-source software should be checked according to its supply chain and manufacturer-reserved backdoors. In addition, even when a program does not use specific vulnerable components directly, an element bundled in some linked package (\mbox{\it{e.g.}} third-party library, module, etc.) may cause problems and then affect others by cascading effects defined as transitive dependency. Furthermore, it is essential to examine source code and its documentation to find vulnerabilities. Although many developers are mindful of secure-code best practices, there is no guarantee that they will follow all guidelines during development phases or integrate them into software processes. Moreover, some problems may still exist in the available code as it is challenging to detect security risks before software deployment~\mbox{\cite{gueye2021decade}}.

New techniques and tools have been created to identify critical software vulnerabilities, such as static and dynamic analyzers. Ponta {\it et al.}~\cite{ponta2018beyond, ponta2020detection} presented the actual research status in open-source software detection methods and tools. They reinforced source-code verification practices by contributing to static analysis and its combination with dynamic approaches to mitigate open-source vulnerabilities. 

Karl Palmskog {\it et al.} \mbox{\cite{palmskog2018picoq}} proposed the tool piCoq to check vulnerabilities in large-scale projects and find failing proofs. This tool can track dependencies between files, definitions, and lemmas and then perform parallel checking of only those files or proofs affected by changes between two project revisions. In the same context, Ruscio {\it et al.}~\mbox{\cite{di2012evoss}} introduced a new software verification tool called EVOSS. It has a fault detector component able to discover inconsistencies in system configuration models. This tool can predict upgrade failures before they can affect a real system. EVOSS has been applied to real Linux distributions (Debian and Ubuntu), which showed improvement in the state-of-the-art package managers for open-source software.

Indeed, security has become a fundamental aspect of system development. Moreover, the way open-source code is created allows anyone to inspect and modify it. Consequently, such a {\it modus operandi} produces rapid prototyping, open access, and community-oriented development. Unfortunately, it also makes the entire scenario riskier regarding intentional and unintentional vulnerabilities, development mistakes, and implementation errors. In addition, weaknesses in open-source libraries can have impressive impacts on the security of elements developed by the software industry due to code reuse and implementation dependencies~\mbox{\cite{plate2015impact}}.

The studies above show the importance of pursuing verification methods for open-source software due to its intrinsic characteristics: collaborative development and code disclosure. It encompasses problems caused by the lack of knowledge during software implementation, given that third-party open-source libraries, components, utilities, and other open-source software are used in a bundle without further analysis. Indeed, most efforts focus on new techniques, and only a few initiatives try to efficiently or effectively tackle the massive amount of associated source code in open-source projects. Due to this clear gap, this paper investigates and tackles security vulnerabilities in large C open-source code bases using a novel methodology that provides an automatic verification framework.

\section{Verifying Security Vulnerabilities in Large Software Systems}
\label{sec:method} 

This section aims to present the proposed verification methodology for large software systems. In addition, the software organization that initially motivated its development is exposed in the next section.

\subsection{The Usual Structure Adopted for Large Open-Source Software Systems}
\label{sec:strucopensource} 

As already stated, the proposed method focuses on verifying large software systems. In addition, there is a special interest in open-source software due to its importance and susceptibility to vulnerabilities. In this context, it is worth mentioning that large open-source software is often organized into multiple files and different folder structures, even referring to external repositories. We can cite some well-known elements as examples, such as PuTTY \mbox{\cite{putty}}, with 175 files, OpenSSH \mbox{\cite{openssh}}, with 276 files, and OpenSSL \mbox{\cite{openssl}}, with $1239$ files. They are regularly used as dependencies in many new projects due to the provision of fundamental communication services, taking advantage of code used in different contexts and domains. Moreover, the number of files that compose a program is usually related to its complexity and constitutes a direct consequence of the chosen design strategy.

Indeed, dividing software into several files can make each of them short enough to be conveniently edited, providing better organization and more straightforward maintenance. For instance, different layers can be placed into different folders, while different components and elements can be split across different files. In addition, source code can be shared with other software implementations in a reuse fashion, e.g., communication buses and network infrastructure. However, this organization also makes verification tasks more complex to manage: each file must be verified, and specific dependencies must be included when checking applications with several elements. In addition, some of these files do not contain the function \textit{main}, which software model checkers usually define as an entry point.

The aspects raised here were considered while developing our verification methodology for large software systems. They were regarded as requirements for any candidate scheme.

\subsection{The Proposed Methodology}
\label{sec:secmeth} 

\begin{figure*}[htb]
	\centering
	\includegraphics[width=17cm]{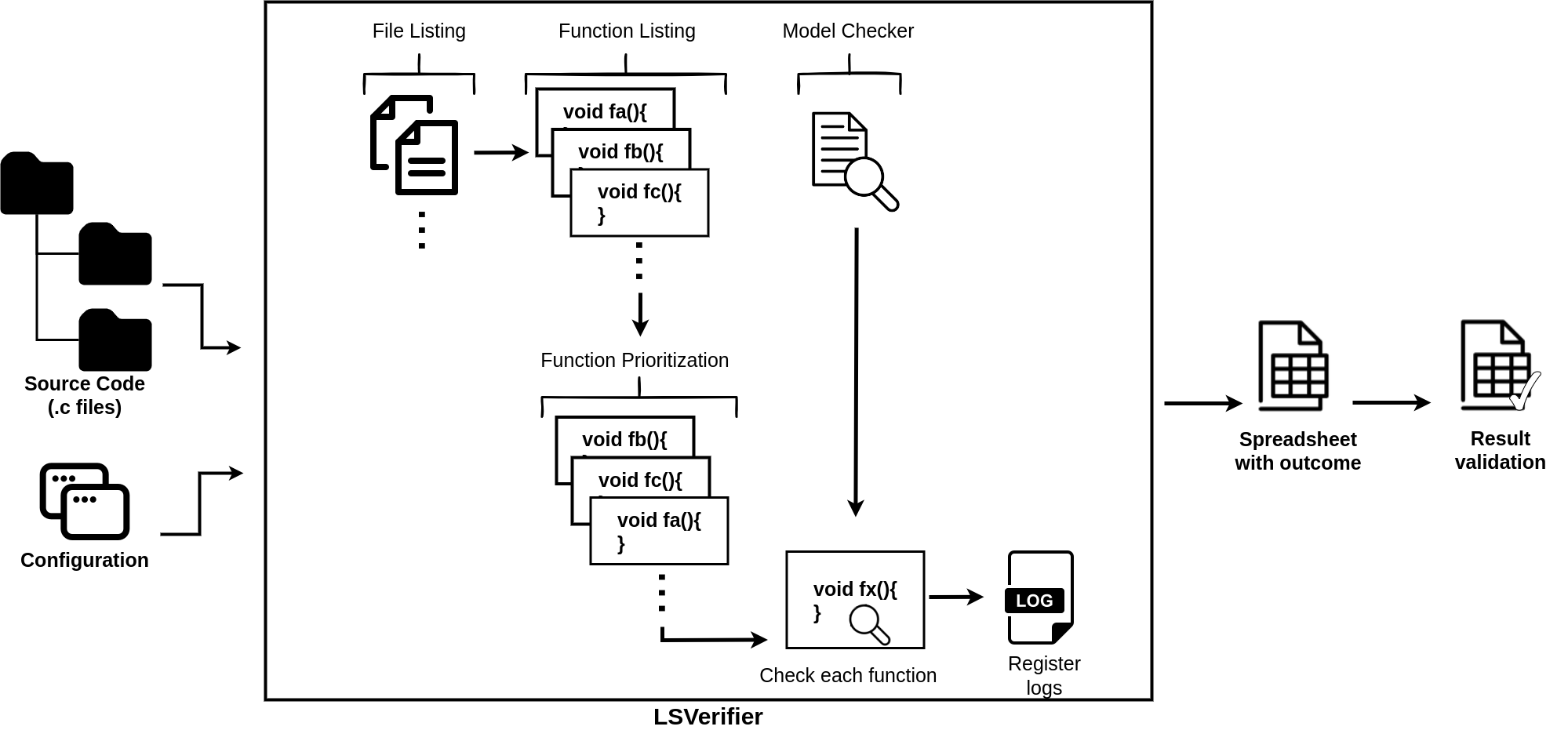}
	\caption{The proposed verification methodology, which begins with the preprocessing of input source-code files, sets the stage for a systematic exploration by the model checker and includes a function-wise prioritization strategy, which streamlines the analysis. Then, it generates a comprehensive report that details all code property violations identified during the verification process. It ends with validating code property violations with a human in the loop.}
	\label{script-new}
\end{figure*}

Inspired by the last section, we have developed a new methodology based on BMC and a prioritized search strategy, whose general idea is illustrated in Figure~\ref{script-new}. It guides an underlying model checker to verify a C program's entire code and can even reach third-party libraries. BMC was chosen as the underlying verification framework due to its performance and flexibility. It can ultimately provide a trade-off between effort (e.g., explored state space and resources) and effectiveness.

First, as illustrated in Figure~\ref{script-new}, the target source-code directory and the necessary configuration, e.g., solver, encoding, and verification methods, are fed. Such information is paramount to match the characteristics of the code to be verified and the goal of the verification process itself. In that regard, one might focus only on a specific type of vulnerability, e.g., overflow. Next, all ``$.c$'' files and all their respective functions and methods are listed. This is an important step that raises the potential locations for vulnerabilities. 

However, in a verification process, users are undoubtedly interested in finding the most dangerous vulnerabilities first, which is inherently linked with the structure of a function or method, e.g., its signature. For instance, it is common sense in the programming community that elements taking raw pointers as parameters are likely more prone to errors than the ones that do not use them~\cite{AlshmranyBBCKMM22}. Consequently, prioritization is also performed, which can be used to drive further test and correction phases to handle more significant problems first. Next, an underlying verifier checks each function according to its priority and the initial configuration. This way, we ensure that the entire source code is evaluated in a prioritized fashion. 

Finally, the associated verification logs are gathered and processed, then used to create a spreadsheet with the related outcomes. Information regarding how files are verified, e.g., way of access and specific vulnerabilities, and how outcomes are displayed, e.g., amount of details, could also be fed. 

A formal description of the proposed method, which can be used for the development of a real implementation of it, is given in Algorithm~\ref{alg:script_algorithm} and is explained as follows. Given a program $P$, in the directory $D$, and the configuration $C$, it first parses the latter. Then it lists the ``$.c$'' files of interest in $D$, which can be performed recursively or be focused on a specific element. As a result, the target files and the input configuration are stored for use in the following steps. 

If a user configures this approach to analyze program functions individually, it will scrutinize each source-code file, generate a comprehensive list $F$ of all functions and methods declared within, and then analyze and reorganize this list according to a prioritization scheme based on the structural elements of each function. Consequently, elements are now sorted from the most to the least priority ones. Otherwise, the analysis occurs based on the normal program flow, using the main function as a starting point with no prioritization scheme. Following that, an underlying BMC checker analyzes each element in $F$, verifying violations such as pointer safety, arithmetic overflow, division by zero, and out-of-bounds arrays. Then, a suitable module generates the respective logs. In that sense, properties to be checked are passed to a controlling script in the argument $configs$. When the associated verification process is concluded, a spreadsheet $V$ with the model checker's outcomes (complete report) is produced, and the compound parsed result of all logs is obtained from a complete execution process.

It is worth noticing that our methodology operates file-wise, which is appropriate and necessary. This way, a file's content, e.g., a complete element, service layer, or interface, is completely verified. Consequently, we can validate source code by investigating building blocks and thus clearing them one by one.

At this point, it is important to clarify the prioritization strategy for functions and methods, i.e., the function {\it prioritized\_functions\_list} in Algorithm \ref{alg:script_algorithm}. Indeed, there can be elements with different signatures, including return and parameter types, which give clues regarding their likelihood to present severe errors. Consequently, depending on them, some components should be evaluated first. For instance, the ones that have pointers or arrays as parameters present an inherent priority. Pointers may be wrongly used throughout a given piece of code and then cause a myriad of problems due to direct memory manipulation. At the same time, arrays may suffer from improper access and incorrect use of their parameters. 

\begin{algorithm}[htb]
	\caption{The proposed verification approach}
	\label{alg:script_algorithm}
	\centering
	\begin{algorithmic}
		\REQUIRE {Program $P$, Directory $D$, Configuration $C$}
		\ENSURE {Verification Outcome V}
		\STATE {$configs \gets get\_configs(C)$}
        \STATE {$function\_analysis \gets extract\_config(configs)$}
		\STATE {$files \gets list\_files(P,D)$}
		\STATE {$num\_files \gets length(files)$}
            \STATE {$log \gets \emptyset$}
		\STATE {$k,l \gets 1$}

		\WHILE {$k \leq num\_files$}
                \IF{$function\_analysis$}
                    \STATE {$F \gets list\_functions(files[k])$}
                    \STATE {$F \gets prioritized\_functions\_list(F)$}
                \ELSE
                    \STATE {$F \gets main\_function$}
                \ENDIF
                    
                \STATE {$num\_functions \gets length(F)$} 

                \WHILE {$l \leq num\_functions$}                
		          \STATE {$log \gets log \cup BMC\_Check(files[k], F[l], configs) $}
		          \STATE {$l \gets l + 1$}
        	\ENDWHILE
                \STATE {$k \gets k + 1$}
	   \ENDWHILE
  
	   \STATE {$V \gets spreedsheat\_create(log)$}
	   \RETURN {$V$}
	\end{algorithmic}
\end{algorithm}  

Moreover, analyzing their bodies is also important to complement the search for potential vulnerabilities. In this context, we first check for dynamic memory allocation, which can potentially cause memory leaks if a developer does not deallocate memory blocks properly. Specifically, we look for the use of memory manipulation functions, e.g., {\it malloc} and {\it free}, to identify this scenario. Next, still within function bodies, we check for the use of asynchronous processing through thread-related calls (e.g., {\it pthread\_create} and {\it pthread\_join} \cite{barney2009posix}), which has the potential to lead to concurrency issues such as race conditions and deadlocks. Finally, the proposed prioritization strategy searches for arithmetic and logical operations such as division and bitwise shifting. These may lead to overflow and odd behavior due to the way they are handled by compilers and converted into low-level instructions.

The proposed prioritization is important for progressive correction procedures. This way, code-fixing phases can focus on severe problems, providing software without the most destructive vulnerabilities. Then, final releases can focus on simpler issues.

Consequently, functions are ranked according to a numerical prioritization scale ranging from grade $5$ (the highest priority) to grade $0$ (the lowest priority) based on the following criteria:
\begin{itemize}
\item the presence of pointers as parameters present the maximum priority and leads to grade $5$;
\item the use of arrays as parameters indicates grade $4$;
\item when dynamic memory allocation is present in a function's body, we tag it with grade $3$;
\item thread manipulation code results in grade $2$;
\item functions with arithmetic operations or bit shifting are classified as grade $1$;
\item the remaining ones are considered low-priority functions and are then tagged with grade $0$.
\end{itemize}

Regarding the prioritization algorithm, it is important to highlight an interesting point: primary functions may be called secondary ones. Indeed, if, when analyzing a function body, there is a call to another function already present in the list of elements to be analyzed, this is removed from it. It is necessary because some elements can be verified more than once since checkers usually follow a function's flow and automatically analyze all elements along it without the need for an explicit request. Consequently, such a removal step inherently leads to a reduction in execution times by avoiding redundant analysis procedures.

Algorithm~\ref{alg:priority} gives a formal description of the function {\it prioritized\_functions\_list} used in Algorithm~\ref{alg:script_algorithm}. It first reads the initial list of functions $F$ and then removes from it functions already called inside other elements ($remove\_from\_list$). Then, if the current function is not part of another context, it assigns different grades for each one, which is done according to the prioritization scheme mentioned earlier. Next, a new list $F^o$ is assembled and returned, which is done with the function $sort\_functions\_by\_priority$. Its content is sorted in descending order, which is based on the newly assigned priorities.

\begin{algorithm}[htb]
	\caption{The proposed prioritization algorithm}
	\label{alg:priority}
	\centering
	\begin{algorithmic}
		\REQUIRE {Function List $F$}
	 	\ENSURE {Ordered Function List $F^o$}
   \STATE{$num\_functions \gets length(F)$}
   		\STATE {$k \gets 1$}
		\WHILE {$k \leq num\_functions$}
              \IF{$\forall f \in F: \, F[k] \supset f$}
              \STATE{$remove\_from\_list(F,f)$}
            \ELSIF{$F[k] \supset pointers$}
            \STATE{$F[k].priority \gets$ grade 5}
            \ELSIF{$F[k] \supset array$}
              \STATE{$F[k].priority \gets$ grade 4}
            \ELSIF{$F[k] \supset (malloc \mid free)$}
              \STATE{$F[k].priority \gets$ grade 3}
            \ELSIF{$F[k] \supset threads$}
              \STATE{$F[k].priority \gets$ grade 2}
            \ELSIF{$F[k] \supset (arithmetic~ operations \mid bit-shift~ operations)$}
              \STATE{$F[k].priority \gets$ grade 1}
            \ELSE
              \STATE{$F[k].priority \gets$ grade 0}
            \ENDIF

            \STATE {$k \gets k+1$}
            \ENDWHILE
            \STATE {$F^o \gets sort\_functions\_by\_priority(F)$}
            \RETURN{$F^o$}
 	\end{algorithmic}
\end{algorithm}  

As already mentioned, the outcomes of a complete verification procedure are exported to a spreadsheet, which contains all property violations found by the underlying BMC checker. This report aims to provide a clear and concise overview of the identified vulnerabilities, including detailed information on each, such as property, file name, function name, and code line where it was detected. This information allows developers to locate and investigate the specific code that may be causing a vulnerability, also estimating its potential impact.

\section{An Implementation of the proposed methodology}
\label{sec:imple}

As our methodology for verifying large software systems had been completely devised, the next logical step was its implementation as a real tool capable of being run and evaluated.

LSVerifier, which is the name chosen for this tool, was implemented as described in Section~\ref{sec:method}, using the programming language Python~\mbox{\cite{python3}}. As the specific underlying BMC checker, ESBMC was chosen, which happened due to its performance in previous instances of the International Competition on Software Verification (SV-Comp)~\cite{Dirk2020,beyer2022progress} (see Section \ref{sec:using-esbmc}). LSVerifier supports all aspects of C11~\mbox{\cite{iso2012}}, the current standard for the C programming language, and detects vulnerabilities in software by simulating a finite prefix of program execution with its possible inputs. Also, an input program is verified by explicitly exploiting interleavings, where one symbolic execution per interleaving is produced. 


By default, LSVerifier can check for various software vulnerabilities corresponding to the ``Top 25'' CWE list by MITRE \cite{topcwes2023}. To do so, it is necessary to specify command-line options, which are linked to our control core's and the ESBMC's options. The vulnerability classes that LSVerifier can be detected include (cf. Section~\ref{sec:background}):

\begin{itemize}
  \item out-of-bounds array access;
  \item illegal pointer dereferences (null dereferencing, out-of-bounds dereferencing, double free, and misaligned memory access);
  \item arithmetic under and overflow;
  \item not a  number (NaN) occurrences in floating-point;
  \item division by zero;
  \item memory leak;
  \item dynamic memory allocation;
  \item atomicity violations at visible assignments.
\end{itemize}

In addition, LSVerifier can prioritize the chosen vulnerability classes according to the parameters configured in the command line. It is also possible to select specific vulnerabilities to be checked, and one can prioritize code exploitation based on specific function types present in source code (see Algorithm~\ref{alg:priority}). The following sections will describe, in detail, the most important implementation modules of the proposed methodology, including their operation and configuration.

\subsection{Software model checking with ESBMC}
\label{sec:using-esbmc}

ESBMC~\cite{Gadelha2019} is a mature bounded model checker that supports the verification of single- and multi-threaded C/C++, Kotlin, Lua, and Solidity programs. It can automatically check predefined safety properties (e.g., memory leaks, pointer safety, array-bounds violations, and overflow) and user-defined software assertions. Besides, ESBMC has been awarded several prizes at SV-Comp~\cite{Dirk2020,beyer2022progress}. There are also ESBMC-based applications used in several domains (e.g., digital controllers~\cite{Cavalcante2020} and photovoltaic systems~\cite{Trindade2019}) as it can check system properties based on description models, which allows capabilities beyond simple basic software checking. 

ESBMC converts input C/C++ programs into a format called GOTO~\cite{HandlingLoop}, replacing all control structures with (conditional) jumps and unrolling loops up to a bound $k$, simplifying a program's representation. This transformation results in an intermediate representation that simplifies the program's structure while preserving its semantics, thereby facilitating the process of model checking and verification. The symbolic execution of this new organization converts software into a static single assignment (SSA) form. 

ESBMC performs various optimizations at the SSA level, including constant propagation to further simplify expressions and instruction slicing to remove unnecessary instructions. Such a simplification process is an important step and also speeds up software verification for some particular applications. First, it removes all instructions after the last assert in an SSA set. Then, it collects all symbols (and their dependent ones) in assertions and removes instructions that do not depend on them. Both phases ensure that unnecessary instructions are ignored in the next step.

The mentioned SSA expressions are then encoded using SMT. If the resulting SMT formula is shown to be satisfiable, a counterexample is presented, which describes the error found; otherwise, there are no errors up to the unwinding bound $k$~\cite{soton416918}.  

To effectively utilize ESBMC, it is essential to configure three key parameters: the choice of SMT solver (options include Z3~\cite{leonardo2008moura}, Boolector~\cite{DBLP:journals/jsat/NiemetzPB14}, Yices~\cite{dutertre2006yices}, MathSAT~\cite{bruttomesso2008mathsat}, or CVC4~\cite{barrett2011cvc4}), the encoding approach (either fixed- or float-point bitvector), and the selected verification technique (\textit{k}-induction, falsification, or incremental BMC). By default, ESBMC employs Boolector as the SMT solver if no specific solver is indicated via the command line. One can set these parameters and other ESBMC's options with the flag \textbf{-e}, as detailed in Table \ref{tab:parameters}.

\subsection{The LSVerifier's Configuration}
\label{sec:SetupConfiguration} 

LSVerifier is a Python software module that allows control via command-line options, which are informed in Table~\ref{tab:parameters}. These options control the following processes: file listing, function verification, outcome display, ESBMC's options, and pointer checking control. The latter refers to a flag responsible for disabling pointer checks during the execution of ESBMC.

\begin{table}[!htb]
  \centering
  \caption{The LSVerifier's Configuration Parameters}
  \begin{tabular}{|p{1.8cm}|p{5.2cm}|}
    \hline
    \rowcolor{black}
    \cellcolor{headercolor}\textcolor{white}{\textbf{Parameter}} & \cellcolor{headercolor}\textcolor{white}{\textbf{Description}} 
    \\ \hline
    -h, --help & Shows the available options\\
    \hline
    -e, --esbmc-parameter & Defines the parameters to be provided to ESBMC.\\
    \hline
    -l file & Provides a file with paths for including header files from dependencies\\
    \hline
    -f, --function & Enables the function verification\\
    \hline
    -fp, --function-prioritized & Enable Prioritized Functions Verification\\
    \hline
    -v, --verbose & Enables the verbose mode\\
    \hline
    -r, --recursive & Enables the recursive verification\\
    \hline
    -d dir & Sets the directory to be verified\\
    \hline
    -p  & Specifies the vulnerability class to be checked\\
    \hline
    -fl file & Specifies a single file to be verified\\
    \hline
    -dp & Disables pointer verification\\
    \hline
  \end{tabular}
  \label{tab:parameters}
\end{table}

Indeed, in a one-by-one fashion, LSVerifier internally verifies all classes of software vulnerabilities that were previously explained. Moreover, with options \mbox{\textbf{-e}} and \mbox{\textbf{-f}}, we can configure a specific property to be explored in C source code.

One may notice that although LSVerifier has been developed to work with ESBMC, it can be integrated with other software model checkers. This can be done by changing arguments to comply with a different set of verification parameters.

\subsubsection{File Listing}
\label{sec:file-listing}

File listing can be performed in three ways: a single file, all elements in the current directory, and a recursive search. To verify a single file, the parameter \textbf{-fl} must be used, informing which element will be handled. This way, it is possible to check all functions of that file or just the main one, the latter being the usual operation mode for most model checkers. 

 Currently, directory and recursive listing are performed with the option \textbf{-r}, using the {\it glob} module, i.e., the UNIX style path name expansion~\cite{glob}. It is employed to find a path name with a specific pattern following the Unix shell's rules. So, for instance, to list all source files written in C, the pattern ``$*.c$'' is used.

\subsubsection{Function Listing and Prioritization} 
\label{sec:func-listing}

LSVerifier lists all functions in a source-code file by using the flag \textbf{-fp}, which guarantees that each of them will be verified, including the one named \textit{main}. It is done with an internal parser that extracts every function's signature and body. 

Specifically, the mentioned parser uses regular expressions to find functions and extract parameter lists from their signatures. These lists provide information and help quickly explore source code in search for vulnerabilities. Next, it copies the entire contents of all functions into specific buffers. These items (i.e., parameter list and function's body), available for every function, are used by the Python module responsible for the prioritization process. Finally, this parser includes a search mechanism that looks for strings representing a function's name inside other functions' bodies (i.e., all content between \{ and \}), avoiding unnecessary calls.

In summary, it is possible to explore source code in a prioritized fashion, which is based on specific function types present in one given file. In the resulting list, functions are ordered according to their information, which readily provides prioritization. 

\subsection{Exporting results}  
\label{sec:exp-results}

After verifying each function in the target files, LSVerifier generates a verification report bearing the corresponding verification outcome. It is encapsulated into a spreadsheet of type comma-separated values (CSV), which allows easy handling and use.

This concise report is employed to analyze an error, understand its root, and then correct either the initial specification or the input software itself. The following items are present in every resulting spreadsheet report written by LSVerifier:
\begin{itemize}
    \item filename;
    \item verification status (e.g., failed);
    \item function name in which the violation was found;
    \item line number in which the function was called;
    \item violation type (e.g., NULL Pointer).
\end{itemize}

If a user wants to check specific results, a companion log file is also provided, which gathers all outputs obtained during an execution.

\subsection{Illustrative Example}
\label{sec:example}

As an illustrative example, this section describes using LSVerifier in a real verification process executed for PuTTY~\cite{putty}, a popular network file transfer application. To check a software piece with several files, as with PuTTY, we have to run it in the directory where its source code is located. 

In C language programming, developers often employ header files that contain declarations of constants, macros, and functions. Compilers typically search for these resources in default directories and folders specified by the compilation command. In our case, their paths must be manually listed in a text file, typically named {\it dep.txt}, which is then passed to LSVerifier during its execution, using the parameter \textbf{-l}, so that it can map them (e.g., third-party libraries). Listing~\ref{list:depfile} illustrates an example of such a file for PuTTY, where each path is included, one per line. 

\begin{listing}[htp]
\caption{The header paths in file {\it dep.txt}} 
\label{list:depfile} 
\footnotesize
\begin{verbatim}
		/usr/include/gtk-3.0/
		/usr/include/glib-2.0/
		/usr/include/pango-1.0/
		/usr/include/cairo/
		/usr/include/gdk-pixbuf-2.0/
		/usr/include/atk-1.0/
            ...
\end{verbatim}
\end{listing}

After listing the associated dependencies, we can run LSVerifier using the parameters described in Section~\ref{sec:SetupConfiguration}. The example in Listing \ref{list:commrun} illustrates a verification process for an entire project, running LSVerifier with arguments configured according to our previous explanation, which was used for PuTTY.

\begin{listing}[htp]
\caption{A command that allows the LSVerifier's function-by-function verification for an entire project} 
\label{list:commrun} 
\footnotesize
\begin{verbatim}
   $ lsverifier -v -r -f -l dep.txt
\end{verbatim}
\end{listing}

\begin{figure}[htb]
	\centering
	\includegraphics[width=8cm]{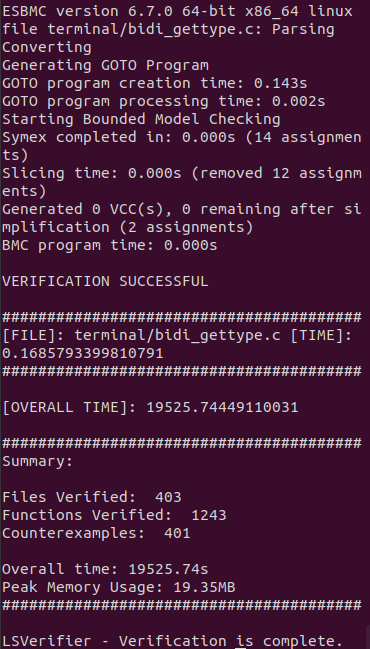}
	\caption{The LSVerifier's output during a verification procedure for PuTTY}
	\label{putty_full}
\end{figure}
\begin{figure*}[htb]
	\centering
	\includegraphics[width=17cm]{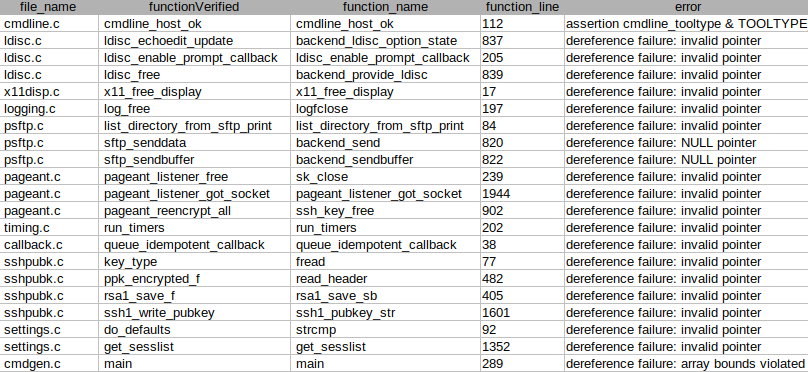}
	\caption{The verification report generated by LSVerifier for PuTTY}
	\label{putty_report}
\end{figure*}

During its execution, LSVerifier checks the properties mentioned in the associated command line, whose progress is informed via logs in the same console. Figure~\ref{putty_full} contains the output of LSVerifier for Putty, while Figure~\ref{putty_report} illustrates the verification report ($.csv$) generated for the same analysis procedure.

\section{Experimental Evaluation}
\label{sec:experimental}

This section presents the experimental evaluation of our approach for verifying large software systems, which focuses on open-source projects developed in the C programming language. In Section~\ref{sec:setup}, we explain the chosen setup. Then, in Sections~\ref{sec:objectives} and~\ref{sec:availability}, we define our experimental goals and show how one can download and reproduce our experiments, which includes scripts, benchmarks, tools, and instructions. Next, in Section \ref{sec:results}, we detail and discuss the obtained results, i.e., identified vulnerabilities, resource allocation, and problem acknowledgment and correction. Finally, in Section~\ref{sec:threats}, we assess threats to the validity of our experiments.   

\subsection{Experimental Setup} 
\label{sec:setup}

All experiments described in this work, using LSVerifier, were performed on a personal computer with an Intel(R) Core(R) i7 CPU 9750H processor and the Ubuntu 20.04 operating system. Moreover, it ran under a clock of $2$.$60$ GHz and used $32$ GB of random access memory (RAM). 

All execution times presented here are CPU times, i.e., only the elapsed periods spent in the allocated CPUs, measured with the Linux tool {\it time} ~\cite{LinuxManual}. LSVerifier used this procedure to compute the total time consumed when verifying software vulnerabilities. Additionally, an approach was devised to assess the peak memory allocation during verification processes. It was achieved using the module {\it tracemalloc} \cite{tracemalloc}, which traces the allocated memory blocks and allows efficient and real-time tracking of memory consumption.

ESBMC v6.7.0 was employed for the verification of C programs. It focused on code robustness regarding accurate pointer utilization, appropriate access to contiguous memory blocks, detection of values leading to variable overflow, and identification of division by zero.

To evaluate our verification methodology, focusing on large software systems, we selected twelve prominent open-source programs written in the C programming language: VLC (version 3.0.18) \cite{VLC}, VIM (version 9.0.1672) \cite{VIM}, Tmux (version 3.3a) \cite{TMUX}, RUFUS (version 4.1) \cite{RUFUS}, OpenSSH (version 9.3) \cite{openssh}, CMake (version 3.27.0-rc4) \cite{CMake}, Netdata (version 1.40.1) \cite{Netdata}, Wireshark (version 4.0.6) \cite{Wireshark}, OpenSSL (version 3.1.1) \cite{openssl}, PuTTY (version 0.78) \cite{putty}, SQLite (version 3.42.0) \cite{sqlite}, and Redis (version 7.0.11) \cite{redis}. They are distributed under open-source licenses, such as the GNU General Public License (GPL), the Apache License, and the Massachusetts Institute of Technology (MIT) License, and more details for each can be found in the respective code repositories. Such programs have been selected due to three main aspects:
\begin{itemize}
    \item large code size;
    \item high importance for the open-source community;
    \item high amount of linked third-party libraries.
\end{itemize}

\subsection{Experimental Objectives}  
\label{sec:objectives} 

The experiments performed here consisted of executing LSVerifier against complex open-source projects listed in Table~\ref{soft_violations}, where different software security properties were checked. Indeed, our experimental evaluation has the following goals.

\begin{tcolorbox}
\begin{enumerate}  

\item[\textbf{EG1}] \textbf{(Automation and Scalability)} Does our approach yield results in a reasonable timeframe and scale efficiently to handle large software systems without requiring manual intervention?

\item[\textbf{EG2}] \textbf{(Practical Use and Effectiveness)} Is our methodology capable of identifying common issues in real software modules that their respective developers corroborate?

\end{enumerate}
\end{tcolorbox} 

EG1 deals with the ability of LSVerifier to guide its underlying checker (i.e., ESBMC) and adapt input code for prompt verification through automated reasoning. EG2, in turn, focuses on its application to practical software and the likelihood of success of approaches that include it, thus finding problems that would be harder to identify without it. 

\subsection{Availability of Data and Tools} 
\label{sec:availability}  

Our execution environment comprises LSVerifier, a precompiled ESBMC binary, and the dataset of open-source projects mentioned in Section \ref{sec:setup}.  LSVerifier and ESBMC\footnote{\url{www.esbmc.org}} are made available in a single package, while the dataset is a compound one. The project's GitHub repository \footnote{\url{https://github.com/janislley/LSVerifier}} contains the Python code, and its Zenodo repository \footnote{\url{https://zenodo.org/records/10077388}} provides all scripts, benchmarks, tools, and instructions to run tests. LSVerifier is publicly available under the MIT License. It can be installed using the tool {\it pip} (Python package installer), with the Linux command in Listing \ref{list:comminstall}.

\begin{listing}[htp]
\caption{The Linux command to install LSVerifier} 
\label{list:comminstall} 
\footnotesize
\begin{verbatim}
     $ pip3 install lsverifier
\end{verbatim}
\end{listing}

ESBMC is publicly available under the terms of the Apache License 2.0. Instructions for building ESBMC are given in its associated file BUILDING (including the description of all dependencies). Finally, ESBMC is a joint project developed by the Federal University of Amazonas in Brazil, the University of Southampton in the United Kingdom, the University of Manchester, also in the United Kingdom, and the University of Stellenbosch in South Africa. 
 
\subsection{Results}
\label{sec:results} 

This section presents the experimental results obtained when LSVerifier was run against the large open-source software systems introduced in Section~\ref{sec:setup}, using our prioritization strategy (see Section \ref{sec:method}). In addition, these results are discussed, providing execution details such as the number of code lines, functions, and issues found for each project, as well as the computational resources employed for each experiment.

\subsubsection{The Evaluation of LSVerifier}  
\label{sec:esbmc-verification}

We have employed LSVerifier to verify the software modules listed in Table~\mbox{\ref{soft_violations}}, where all C files were individually analyzed, checking each function. We have also used flags for log plotting and provision of the ESBMC's configuration.  

This verification process resulted in several violated properties, as shown in Table~\ref{soft_violations}, which were mostly related to \textit{dereference failure} as explained in Definition \ref{NullPointerDerefence}. This table includes the modules' names and versions, the number of violated properties, the amount of ``$.c$'' files, the number of external inclusions (``.h'' files), the amount of source-code lines, the number of verified functions, the elapsed execution times, and the maximum memory usage. All the essential vulnerabilities, according to CWE, i.e., pointer dereference, division by zero, dynamic object violation, and array-bounds violation, were identified during our analysis.   

\begin{table*}[htbp]
    \centering
    \caption{Software vulnerabilities detected by LSVerifier, correlating the number of code-property violations with the amount of C code files, external libraries, lines of code, functions verified, maximum necessary memory, and verification time}
    \begin{tabular}{|l|c|c|c|c|c|c|c|c|c|c|}
    \hline
    \cellcolor{headercolor}\textcolor{white}{\textbf{\makecell[tc]{Software \\ Project }}} & \cellcolor{headercolor}\cellcolor{headercolor}\cellcolor{headercolor}\cellcolor{headercolor}\cellcolor{headercolor}\textcolor{white}{\textbf{\makecell[tc]{Software  \\ Version}}} &   \cellcolor{headercolor}\cellcolor{headercolor}\cellcolor{headercolor}\cellcolor{headercolor}\textcolor{white}{\textbf{\makecell[tc]{Property \\ Violated}}} & \cellcolor{headercolor}\cellcolor{headercolor}\cellcolor{headercolor}\textcolor{white}{\textbf{\makecell[tc]{Files}}} & 
    \cellcolor{headercolor}\cellcolor{headercolor}\textcolor{white}{\textbf{\makecell[tc]{External \\ Includes}}} & 
    \cellcolor{headercolor}\textcolor{white}{\textbf{\makecell[tc]{Source-code \\ Lines}}} & 
    \cellcolor{headercolor}\textcolor{white}{\textbf{\makecell[tc]{Functions \\ Verified}}} & 
    \cellcolor{headercolor}\textcolor{white}{\textbf{\makecell[tc]{Verification \\ Time}}} & 
    \cellcolor{headercolor}\textcolor{white}{\textbf{\makecell[tc]{Memory \\ Usage}}} \\ 
    \hline
    VLC		  & 3.0.18     & 72   & 1171 
    & 289 & 421840 & 13709  & 1033.79s  & 20.09MB  \\ \hline
    VIM		  & 9.0.1672   & 110  & 188  
    & 95  & 366775 & 9611   & 554.56s   & 39.83MB  \\ \hline
    TMUX	  & 3.3a       & 1788 & 179  
    & 445 & 61004  & 2168   & 52218.45s & 43.12MB  \\ \hline
    RUFUS	  & 4.1        & 576  & 144  
    & 108 & 56278  & 1615   & 283.95s   & 6.06MB   \\ \hline
    OpenSSH	  & 9.3        & 338  & 290  
    & 63  & 109791 & 3183   & 873.27s   & 42.58MB  \\ \hline
    Cmake	  & 3.27.0-rc4 & 552  & 1516 
    & 1030& 324760 & 11279  & 934.21s   & 37.07MB  \\ \hline
    Netdata	  & 1.40.1     & 1318 & 307  
    & 160 & 312530 & 7352   & 51471.27s & 129.09MB \\ \hline
    Wireshark & 4.0.6      & 2141 & 2330 
    & 77   & 4177163& 121567 & 59952.39s & 391.44MB \\ \hline
    OpenSSL   & 3.1.1      & 3140 & 1575 
    & 616 & 491632 & 17168  & 6046.63s  & 53.34MB  \\ \hline
    PuTTY	  & 0.78       & 2472 & 403  
    & 153 & 127282 & 5310   & 66210.32s & 58.54MB  \\ \hline
    SQLite    & 3.42.0     & 3265 & 340  
    & 609 & 258382 & 8911   & 2493.75s  & 33.22MB  \\ \hline
    Redis	  & 7.0.11     & 187  & 418  
    & 556 & 170673 & 8211   & 727.76s   & 46.57MB  \\ \hline
    \end{tabular}
    \label{soft_violations}
\end{table*} 

LSVerifier was able to detect potential vulnerabilities in every tested software project. The lowest number of violations occurred with VLC, which is not necessarily a surprise, considering its development time ($20$ years) and its importance as one of the leading open-source media players currently available. Although VLC has $13709$ functions split throughout $1171$ files, with $421840$ lines of code, its verification time is relatively short (above $17$ minutes) when compared with others that are available in the same table. For instance, Tmux, which comprises $2168$ functions distributed across $179$ files, totaling $61004$ lines of code, required approximately $14.5$ hours for verification. 

When applied to other large open-source software such as SQLite, OpenSSL, Putty, and Wireshark, all with more than $120000$ code lines, LSVerifier identified many property violations: $3265$, $3140$, $2472$, and $2141$, respectively, as shown in Table~\ref{soft_violations}. These high numbers of vulnerabilities are, primarily, a consequence of the use of multiple superficially-checked third-party libraries, which led to a lot of header-file inclusions (i.e., ``.h'' files) in source code: $609$ for SQLite, $616$ for OpenSSL, $153$  for PuTTy, and $77$ for Wireshark. Such behavior leads to problems that may not have been considered in unit tests or tackled during testing rounds.  

The highest peak memory usage was observed with Wireshark, which presents a complex code organization and the highest number of functions, files, and lines of code. This project presents $77$ header-file inclusions, as already mentioned, and undertakes the complex task of network traffic analysis. When used as a tool, running on Linux, Wireshark usually requires more than $500$ MB of RAM, which already hints at its high resource demands. In addition, Wireshark imposes no limit on the number of packets it can handle,  which creates a rich state space to be explored by LSVerifier. Netdata is another project that requires high memory usage for analysis, which performs media decoding and presentation. This project is organized in $312530$ lines of code and includes $160$ different external header files.

In terms of verification duration, PuTTY required the longest time, which was unexpected given its relatively small codebase, consisting of $127282$ lines of code, $5310$ functions, and $153$ includes. Nevertheless, PuTTY executes complex operations, which involve encrypted communications and various protocols. Wireshark accounted for the second-longest verification time despite having a significantly larger code volume. This suggests that the analysis complexity is not solely dependent on the size of a program or the number of its header-file inclusions but rather on the intricacy of its programming and structures, which its dependencies may also influence.

LSVerifier was able to check programs with sizes ranging from $144$ files, $1615$ functions, and $56268$ code lines, which is the case of RUFUS, to $2330$ files, $121567$ functions, and $4177163$ code lines, which is the case of Wireshark. The obtained figures clearly show scalability capacity. Furthermore, the peak memory usage ranged from $6.06$ to $391.44$ MB of RAM, which is an acceptable amount given the typical hardware capabilities of modern personal computers. This indicates that LSVerifier can maintain low memory requirements, even with large software. It also significantly differs from other recent verification tools based on model checking, focusing primarily on execution speed and CPU usage~\cite{mann2021pono,lange2020ic3,chenoy2021c}.

Putty required $18$ hours to be verified, representing the longest verification procedure. In other words, software modules with hundreds of thousands or even millions of lines of code can be evaluated in less than one day in a completely automated manner, which one can even trigger after a code-delivery meeting for nightly execution. 

\begin{tcolorbox}
Consequently, the above aspects reinforce the scalability properties of LSVerifier and integrally answer \textbf{EG1}: the proposed approach produces results in a reasonable period, and it can be applied to small and very large software systems indistinctly.
\end{tcolorbox}

Although Table \ref{soft_violations} shows the total amount of violations, it is also interesting to evaluate the prevalence of distinct vulnerability classes. Table \mbox{\ref{soft_violations_analysis}} presents the same verification results in Table \ref{soft_violations} but now categorized into eleven different types of property vulnerabilities detected by LSVerifier, following the ESBMC's nomenclature:

\begin{itemize}
\item \textbf{Invalid Pointer (IP)}, which corresponds to null pointer dereferences, as described in Definition \ref{InvalidPointer};

\item \textbf{Array Bounds Violated (ABV)}, \textbf{Array Lower Bound (ALB)}, and \textbf{Array Upper Bound (AUB)}, which are specific cases where an array is accessed beyond its allocated boundaries, as detailed in Definition \ref{outofbounds}, and are intrinsically linked to buffer overflows, where data overruns the set limits of a buffer, as described in Definition \ref{BufferOverflow};

\item \textbf{Same Object Violation (SOV)}, which happens when pointers are compared in violation of the "same object" rule, i.e., C language allows the comparison of pointers using relational operators but imposes restrictions on their use when operands are pointers referring to the same address, as described in Definition \ref{sameobjectvio};

\item \textbf{Invalid Pointer Freed (IPF)}, which occurs when an uninitialized or invalid pointer is released using the function {\it free}, as described in Definition \ref{InvalidPointer}, and is similar but not the same as IP, focusing on memory deallocation;


\item \textbf{Invalidated Dynamic Object (IDO)}, which corresponds to objects that become invalidated, often as a result of deallocation, can arise from various sources, mainly pointer problems, 
and is further elaborated in Definitions \ref{InvalidPointer}, \ref{NullPointerDerefence}, and \ref{DoubleFree}; 

\item \textbf{Null Pointer (NP)}, which involves the inappropriate use of null pointers, typically by dereferencing them, as described in Definition \ref{NullPointerDerefence}, and is specifically about null pointers (IP and IPF may include other types of invalid pointers);

\item \textbf{Division by Zero (DZ)}, which refers to cases where a number is divided by zero and causes undefined behavior, as described in Definition \ref{Dividebyzero}, and is also a specific type of arithmetic operation that can lead to undefined behavior, as described in Definition \ref{ArithmeticOverflow};

\item \textbf{Assertion Failure (AF)}, which happens when a condition passed to the function \textit{assert} is not met, indicating an error or unexpected behavior, as described in Definition \ref{AssertionFailure};

\item \textbf{Access to Object Out of Bounds (AOOB)}, which is a more general term that could refer to accessing any object (e.g., strings and linked lists), not just arrays, beyond their allocated boundaries, is primarily described in Definition \ref{ImproperBufferAccess}, and is also associated with several other CWEs;
\end{itemize}

\begin{table*}[!htb]
\centering
\scriptsize
\caption{Software violations found, using the chosen dataset, with LSVerifier and split into the $11$ categories recognized by ESBMC}.
\begin{threeparttable}
\centering
\begin{tabular}{|l|c|c|c|c|c|c|c|c|c|c|c|c|c|c|}
    \hline  
    \rowcolor{black}
    \cellcolor{headercolor}\textcolor{white}{ \textbf{Software}} &   
    \cellcolor{headercolor}\textcolor{white}{ \textbf{Source}}   &  
    \cellcolor{headercolor}\textcolor{white}{ \textbf{Header Paths}} & 
    \cellcolor{headercolor}\textcolor{white}{ \textbf{IP}}  & 
    \cellcolor{headercolor}\textcolor{white}{ \textbf{ABV}}  & 
    \cellcolor{headercolor}\textcolor{white}{ \textbf{ALB}} & 
    \cellcolor{headercolor}\textcolor{white}{ \textbf{AUB}}  & 
    \cellcolor{headercolor}\textcolor{white}{ \textbf{SOV}} & 
    \cellcolor{headercolor}\textcolor{white}{ \textbf{IPF}} & 
    \cellcolor{headercolor}\textcolor{white}{ \textbf{IDO}}  & 
    \cellcolor{headercolor}\textcolor{white}{ \textbf{NP}}   & 
    \cellcolor{headercolor}\textcolor{white}{ \textbf{DZ}} & 
    \cellcolor{headercolor}\textcolor{white}{ \textbf{AF}} & 
    \cellcolor{headercolor}\textcolor{white}{ \textbf{AOOB}} 
    \\ \hline
    VLC		  & 421840 & 9395 & 57	  & 2 	& 0		& 0		& 0   	& 2 	& 0	  	& 1 	& 0 	& 10  & 0   \\ \hline
    VIM       & 366775 & 443 & 100   & 3	& 1   	& 2	  	& 0		& 2		& 0		& 2	    & 0	    & 0   & 0	\\ \hline
    TMUX      & 61004 & 1034 & 1725  & 0	& 12   	& 9	  	& 0		& 21	& 0		& 20    & 0	    & 1	  & 0   \\ \hline
    RUFUS 	  & 56278 & 1453 & 513   & 0	& 0   	& 4	  	& 4		& 6	    & 0 	& 20   	& 29	& 0   & 0   \\ \hline
    OpenSSH   & 109791 & 3919 & 311   & 3	& 4   	& 0    	& 0		& 4		& 1	    & 10    & 5	    & 0   & 0	\\ \hline
    Cmake 	  & 324760 & 7710 & 481   & 28	& 5   	& 2    	& 18	& 7	    & 0	    & 6	    & 2	    & 3	  & 0   \\ \hline
    Netdata   & 312530 & 1516 & 1045  & 1	& 12  	& 5     & 3 	& 5	    & 0	   	& 212	& 35	& 0	  & 0   \\ \hline
    Wireshark & 4177163 & 19513 & 1940  & 20	& 12  	& 17    & 35	& 2	    & 0		& 77    & 5	    & 27  & 6   \\ \hline
    OpenSSL   & 491632 & 9892 & 2753  & 77	& 98  	& 22    & 10	& 7	    & 2		& 131	& 11	& 29  & 0   \\ \hline
    Putty 	  & 127282 & 2041 & 1996  & 8	& 25  	& 26    & 4		& 6	    & 0  	& 56	& 14	& 337 & 0   \\ \hline
    SQLite 	  & 258382 & 1224 & 2254  & 36	& 15  	& 37    & 16	& 9	    & 0  	& 540	& 29	& 326 & 3   \\ \hline
    Redis 	  & 170673 & 2555 & 150   & 3	& 9  	& 3     & 11	& 3	    & 0  	& 7 	& 0 	& 0   & 1   \\ \hline
\end{tabular}
\label{soft_violations_analysis}
\end{threeparttable}
\end{table*} 

Some identified vulnerabilities were related to conditions that usually lead to memory corruption or crashes, e.g., accessing invalid pointers or out-of-bounds arrays. Although these definitions seem to present new and specific conditions, all of them can be traced back to the basic vulnerabilities in Section \ref{sec:background} as explicitly shown. The comprehensive list of CWEs supported by LSVerifier can be found in Table \ref{cwelist}.

When analyzing Table \ref{soft_violations_analysis}, we can observe that the most prevalent vulnerabilities were the ones related to violations involving invalid pointers, i.e., IP and NP. Indeed, this sheds some light on memory corruption as a critical issue in C source code. Our results showed that overstepping bounds caused most identified pointer safety violations. In such scenarios, pointers were initialized with memory blocks allocated dynamically, and programming mistakes led to out-of-bounds errors. The highest number of pointer violations (IP and NP) was found in OpenSSL, which provides implementations for the protocols' secure sockets layer (SSL) and transport layer security (TSL), with basic encryption capacity. Indeed, due to data-block encryption, there is a lot of memory allocation and pointer manipulation (e.g., using function pointers), which explains the behavior and the associated results. 

OpenSSL presented the highest number of bounds violations too, i.e., ABV, ALB, and AUB ($197$ occurrences). Again, it is also closely related to manipulating memory buffers, which are processed for subsequent use \cite{opensslOOB}. Mishandling these buffers can lead to out-of-bounds errors when data is read or written beyond the allocated memory.

LSVerifier found $35$ SOV and $6$ AOOB occurrences in Wireshark. The latter deals with analyzing network packets, where pointers are extensively used to verify data fields in a unique comparison step, often involving strings and other objects. Consequently, it is expected that LSVerifier would detect operations comparing pointers that refer to the same address and access to objects.

Most IPF events were found in Tmux, a terminal multiplexer developed by various developers. Besides, it relies on external libraries while interacting with an operating system's APIs. During these operations, the resulting tasks undoubtedly require new memory blocks, which must later be freed. These complex routines can explain the obtained results.

SQLite is an engine for accessing structured query language (SQL) databases widely used in embedded device projects, which presented $540$ errors in the NP category. Indeed, SQLite often deals with operation interleaving, corrupted databases, and statements, which may result in NP occurrences \cite{sqlitenp1,sqlitenp2}. In other words, its routine tasks are closely related to this kind of fault, which should inherently lead to more careful coding and testing processes. Moreover, this aspect is challenging in the C language and requires expertise.

Netdata presented $35$ DZ errors. It is a project designed to collect server metrics, assisting system administrators to take proactive measures. Due to its capability to retrieve statistics from different sources along with their associated data (e.g., timestamps), with subsequent computation involving these via external plugins, there is a high risk related to this kind of fault \cite{netdatamectrics}.

When analyzed by LSVerifier, PuTTY presented $337$ AF occurrences. It supports network protocols and manages user inputs by applying concurrency through multiple threads. Consequently, issues related to data validation using assertions, which are usual in this task, can result in assertion failures.

It is worth noticing that some out-of-bounds errors occurred in scenarios where functions wrongly read data from heap-allocated memory. It can corrupt memory or induce unpredictable behavior. For instance, it is possible to exploit a bounds violation to write arbitrary code into some specific positions of memory and afterward execute that same code, which may result in a loss of control over the specific process, compromising the whole software module. OpenSSL, SQLite, and Putty presented the highest hits in the out-of-bounds category, with $197$, $88$, and $59$ failures, respectively. These projects have a significant number of code lines and have been maintained and modified for a long time, which can explain the number of failures for them.

As a general comment, developers must be aware of potential memory management issues so they can take measures to prevent them, for instance, implementing defensive programming practices such as boundary checking on memory access operations. By prioritizing secure memory management practices, developers can help prevent serious software vulnerabilities in their projects.

Finally, we were able to find DZ vulnerabilities in VLC \cite{vlc_issue_1}, RUFUS \cite{RUFUS_issue_1, RUFUS_issue_2, RUFUS_issue_3}, OpenSSH \cite{openssh_issue_1, openssh_issue_2}, and Netdata \cite{netdata_issue_1, netdata_issue_2}. This kind of issue usually occurs when a parameter within a division operation determines the size of the variable to be created before executing an operation. Often, there is no check to assure that the divisor is strictly positive, resulting in an integer overflow bug and a division by zero. The highest threat from this vulnerability regards the system's integrity. Moreover, an attacker can easily disrupt its operability by sending an invalid interval value.

By leveraging LSVerifier's comprehensive detection capabilities, we can create a more robust and resilient tool for software vulnerability management and prevention. The enhancement in security achieved through this approach not only mitigates risks but also promotes a more reliable, efficient, and secure software ecosystem. 

Table \ref{cwelist} relates the CWE identifiers that MITRE has assigned to the specified types of vulnerabilities that LSVerifier can identify. 
It highlights how comprehensive the verification performed by LSVerifier is, given that some categories individually gather many CWE identifiers. 

These vulnerabilities show the importance of the rigorous software testing and validation performed by LSVerifier. Static analysis tools, dynamic analysis, and formal verification methods are crucial in identifying and mitigating these risks early in the development lifecycle. By addressing these vulnerabilities, developers can ensure the robustness, reliability, and security of their software applications.

\begin{table}[!htb]
\centering
\caption{The vulnerabilities identified by LSVerifier along with the corresponding CWE numbers for each property violation detected}
    \begin{tabular}{|p{4.3cm}|p{3cm}|}
    \hline 
    \rowcolor{black}
    \cellcolor{headercolor}\textcolor{white}{\textbf{Vulnerability type}}  & 
    \cellcolor{headercolor}\textcolor{white}{\textbf{CWE numbers}} 
    \\ \hline
    Invalid pointer (IP)  & CWE-416, CWE-476, CWE-690, CWE-822, CWE-824, CWE-908   \\ \hline
    Array Bounds Violated (ABV), Array Lower Bound (ALB), Array Upper Bound (AUB) & CWE-20, CWE-119, CWE-120, CWE-121, CWE-125, CWE-129, CWE-131, CWE-193, CWE-628, CWE-676, CWE-754, CWE-755, CWE-787, CWE-788  \\ \hline
    Same Object Violation (SOV) & CWE-125, CWE-170, CWE-193, CWE-466, CWE-469, CWE-682, CWE-787 \\ \hline
    Invalid Pointer Freed & CWE-415, CWE-416, CWE-459, CWE-590, CWE-761, CWE-825 \\ \hline
    Invalidated Dynamic Object (IDO) & CWE-415, CWE-416, CWE-476, CWE-664, CWE-789    \\ \hline
    NULL pointer dereference  & CWE-391, CWE-476                        \\ \hline         
    Division by zero & CWE-369         			                        \\ \hline
    Assertion violation & CWE-190, CWE-191, CWE-389, CWE-478, CWE-571, CWE-569, CWE-617, CWE-670, CWE-680, CWE-681, CWE-682, CWE-685, CWE-754  \\ \hline
    Access to Object Out of Bounds (AOOB) & CWE-119, CWE-125, CWE-170, CWE-193, CWE-466, CWE-682, CWE-787, CWE-823   \\ \hline
\end{tabular}
\label{cwelist}
\end{table} 

In summary, LSVerifier can find usual problems in real software modules, which are widely known and understood by the development community. This aspect also leads to a prompt explanation and then faster identification and correction. For instance, a developer can quickly understand an IP or NP occurrence, also mentioned in the LSVerifier's final report (spreadsheet). 

\begin{tcolorbox}
Consequently, this capability begins answering \textbf{EG2} because usual and easily understandable problems are found. 
\end{tcolorbox}

However, a deeper discussion regarding problem confirmation is still missing, which will be tackled in the next section.

\subsubsection{Discussion Regarding the Violated Properties}  
\label{sec:putty-prop-analysis}

\begin{table*}[htb]
    \centering
    \caption{The issues, on code property violations, reported to the software project repositories for confirmation of the associated software vulnerabilities by the projects' developers}
    \begin{tabular}{|l|c|c|c|c|}
    \hline 
    \rowcolor{black}
    \cellcolor{headercolor}\textcolor{white}{\textbf{Software}} & 
    \cellcolor{headercolor}\textcolor{white}{\textbf{Issues opened}} & 
    \cellcolor{headercolor}\textcolor{white}{\textbf{Issues confirmed}} & 
    \cellcolor{headercolor}\textcolor{white}{\textbf{Issues fixed}}          
    \\ \hline
    VLC         & 1 \cite{vlc_issue_1} & 1 & 1  \\ \hline
    VIM         & 1 \cite{vim_issue_1} & 0 & 0  \\ \hline
    TMUX        & 1 \cite{TMUX_issue_1}  & 0 & 0                    \\ \hline
    RUFUS       & 2 \cite{RUFUS_issue_2, RUFUS_issue_3} & 2 & 1 \\ \hline
    OpenSSH	    & 2 \cite{openssh_issue_1, openssh_issue_2} & 0 & 0 \\ \hline
    CMake       & 1 \cite{cmake_issue_1} & 1 & 1 \\ \hline
    Netdata	    & 2 \cite{netdata_issue_1, netdata_issue_2} & 0 & 0 \\ \hline
    Wireshark   & 1 \cite{wireshark_issue_1} & 1 & 1 	\\ \hline
    OpenSSL	    & 1 \cite{openssl_issue_1} & 1 & 0 	\\ \hline
    Putty       & 1 (E-mail) & 0 & 0  					\\ \hline
    SQLite	    & 2 \cite{SQLite_issue_1, SQLite_issue_2} & 1 & 0 	\\ \hline
    Redis       & 2 \cite{redis_issue_1, redis_issue_2} & 1 & 0  					\\ \hline
    \end{tabular}
    \label{properties_analysis}
\end{table*} 

It is worth mentioning that the complete chain for software verification involves vulnerability identification, vulnerability confirmation, code analysis, and code repair (e.g., patch application and merge requests to a repository). However, that was not possible for all programs mentioned in Section \ref{sec:setup} due to the infrastructure available for each project, the long time necessary for that, or the availability of their respective developers. Anyway, some of the issues found here were reported within the code repositories of VLC, VIM, RUFUS, OpenSSH, CMake, Netdata, Wireshark, OpenSSL, Putty, and Redis via specific tools or e-mail. In response to the confirmed software vulnerabilities, triggered by fault reports generated by our tool, remedial patches were subsequently applied. 

The issues found and reported in the context of this work were presented based on the counterexample traces provided by LSVerifier. They were also discussed with the respective developers and maintainers, who confirmed some of them and classified others as false positives. Table \mbox{\ref{properties_analysis}} shows the most critical property violations and the corresponding bugs reported via GitHub or e-mail. Although the number of issues informed in the respective repositories may seem low, it is worth mentioning that each takes a long time for proper registration, discussion, and correction, which may even span several months. Consequently, we chose only the most significant ones. Our prioritization strategy directly favored this procedure.  

During the analysis of RUFUS, we were able to identify property violations such as array bounds, division by zero, and invalid pointers. To check these problems with the respective software developers, we have registered $3$ issues \mbox{\cite{RUFUS_issue_1, RUFUS_issue_2, RUFUS_issue_3}}, which were caused by imported libraries and are detailed as follows.

Thus far, we have got one bug fix for $tiny-regex-c$ to handle an out-of-bounds violation \cite{RUFUS_issue_2}, which is related to CWE-787. In Listing \ref{list:ruf1}, which shows real log information, one can notice that this problem happened in file {\it re.c}, one of the two that this dependence presents (the other is {\it re.h}). The value retrieved from the attribute {\it type} of the object {\it pattern} can go beyond the maximum length of the array {\it types}.

\begin{listing}[htp]
\caption{Bounds violation in tiny-regex-c used by Rufus} 
\label{list:ruf1} 
\footnotesize
\begin{verbatim}
Building error trace

Counterexample:

State 5 file re.c line 269 
In function re_print thread 0
-----------------------------------------------
Violated property:
file re.c line 269 function re_print
array bounds violated: 
array `types' upper bound
(signed long int)(pattern + 
(signed long int)i)->type < 17
    
VERIFICATION FAILED
\end{verbatim}
\end{listing}

To avoid such a condition, a specific check was added, which is shown in the code excerpt present in Listing \ref{list:ruf2}. It assures that the index passed to {\it types} is below its limit, i.e., {\it NOT\_WHITESPACE}, as declared in an enumeration. As one can see, this is a simple measure that should be always adopted as common practice. However, it also reveals the careless coding performed by many developers. 

\begin{listing}[htp]
\caption{Corrected C program for tiny-regex-c} 
\label{list:ruf2} 
\footnotesize
\begin{verbatim}
...
if (pattern[i].type <= NOT_WHITESPACE)
    printf("type: %s", 
        types[pattern[i].type]);
else
    printf("invalid type: %d", 
        pattern[i].type);
...
\end{verbatim}
\end{listing}

The second issue is a division by zero related to CWE-369. It was found in the library {\it ext2fs} and discussed with its developers \cite{RUFUS_issue_3}. They acknowledged that if a {\it hashmap} with size $0$ is created, it could cause a program crash. However, this is considered a bug at the application level based on the assumption that such an operation should not be performed. Its identification is shown in Listing \ref{list:ruf3}.

\begin{listing}[htp]
\caption{Divison by zero in ext2fs library used by RUFUS} 
\label{list:ruf3} 
\footnotesize
\begin{verbatim}
Building error trace

Counterexample:

State 4 file hashmap.c line 51 
In function ext2fs_hashmap_add thread 0
-----------------------------------------------
Violated property:
file hashmap.c line 51 
in function ext2fs_hashmap_add
division by zero
h->size != 0
    
VERIFICATION FAILED
\end{verbatim}
\end{listing}

This argument holds from a purely functional standpoint. However, from a security perspective, any unexpected behavior, including crashes, should be considered a potential vulnerability. It is necessary to treat such conditions as potential threats until they are comprehensively analyzed and discarded or properly registered. To address this, a check should be implemented to confirm that {\it h$\rightarrow$size} is not $0$ before the modulo operation, as shown in the code excerpt in Listing \ref{list:ruf4}.

\begin{listing}[htp]
\caption{Corrected C program for ext2fs-c} 
\label{list:ruf4} 
\footnotesize
\begin{verbatim}
    ...
int ext2fs_hashmap_add(
    struct ext2fs_hashmap *h,
    void *data, const void *key, 
    size_t key_len)
{
    // Check if h->size is zero
    if (h->size == 0) {
        // Handle the error
    }

    uint32_t hash = 
        h->hash(key, key_len) % h->size;
    ...
\end{verbatim}
\end{listing}

Again, this is a simple correction that should be a coding rule and could prevent serious problems. In summary, we advocate that even if some vulnerability is not likely to happen due to a given program's structure, it must be handled. This way, even intended bad coding can be reduced, aiming at exploiting known vulnerabilities can be mitigated.




When checking the violated properties for VLC \cite{vlc_issue_1}, some memory-safety vulnerabilities found during our experiments were reported via email. After their analyses, double-free errors were confirmed, a type of vulnerability related to CWE-415. This error occurs when software modules free a memory allocation twice, and doing so can lead to the modification of unexpected memory blocks, even resulting in a system crash or potentially allowing an attacker to execute arbitrary code. Specifically, this fix removed a deprecated Linux framebuffer \cite{framebuffer} plugin. Indeed, the Linux fbdev \cite{fbdev} subsystem has been deprecated for more than a decade as there are better options currently available. 

We have also reported another issue caused by a third-party library in OpenSSL \mbox{\cite{openssl_issue_1}}. The developers confirmed that an invalid pointer is dereferenced, which may likely cause a crash in the caller, but they do not consider this to be a vulnerability as many OpenSSL's APIs crash if a null pointer is passed to them. Here, we clearly have a bad practice. Although a problem was found and confirmed, developers often state that a particular condition may never happen as a specific function or method is never invoked the way it should be to provoke it. If an attacker is aware of that and manipulates parameters or even code to create that unfeasible scenario, the problem found here will really happen and may even cause severe loss. Consequently, we also need to encourage a change of behavior, where any problem is handled properly and treated as a priority. Anyway, such a condition should be monitored as it can be a source for an attack resulting in a system crash.

Regarding CMake, most of its problems are related to third-party libraries. Specifically, our analysis of its property violations revealed an important issue: a confirmed dereference failure caused by an invalid pointer. This error was fixed in the Cmake repository in function $cm\_utf8\_decode\_character(...)$ and was caused by an empty input range local variable resulting in an invalid pointer \mbox{\cite{cmake_issue_1}}, being related to CWE-824. Additionally, CMake employs third-party libraries such as $cmbzip2$ and $cmzstd$, each with its own upstream software source, and our analysis uncovered issues in both. Specifically, we identified problems caused by invalid pointers, which lead to memory corruption and present an opportunity for arbitrary code execution. Furthermore, we were unable to locate any open-source repositories for $cmbzip2$ and $cmzstd$ where we could report these potential vulnerabilities. Here, we highlight another aspect: who is responsible for a given open-source module? Sometimes, it is difficult to answer such a question. As alternatives, it could be removed or even treated as a responsibility of a given project, which may even generate a fork. It can then be later published and regularly maintained.



    

LSVerifier identified issues in Wireshark \mbox{\cite{wireshark_issue_1}} related to array access (bounds), invalid pointer, and null pointer, which are associated to CWE-125, CWE-824, and CWE-476. All these issues were identified in the libraries CMake and network programming language (NPL), which are project dependencies, where dereference failures occurred due to out-of-bounds access and NP occurrences. The third-party library NPL is an ongoing project that has not been prioritized in recent times. The last significant update to this module was approximately nine years ago, with the latest reference in the commit logs dating back eight years. To maintain the robustness and security of Wireshark, the development team opted to remove this module.

As of now, we have not been able to validate the issues associated with VIM, Netdata, and OpenSSH. Indeed, some were classified as false positives \cite{vim_issue_1, openssh_issue_1, openssh_issue_2, netdata_issue_2} by developers, and others are still under discussion \cite{netdata_issue_1}. However, we need to mention that the related problems do exist and suffer from the same problematic practices previously detailed in the OpenSSL context.

Another issue involving multiple instances of invalid pointer dereference was reported for Putty. Numerous memory-related property violations, as detailed in Table \mbox{\ref{properties_analysis}}, were identified. However, since the only communication channel available for Putty is email, we have not received any feedback yet. Consequently, a given weakness may last long and cause a lot of damage.

Regarding TMUX, we have opened one issue \cite{TMUX_issue_1} related to two different violations: null pointer dereference (CWE-476) and array out-of-bounds (CWE-125). The Null pointer dereference occurs in the function $cmd\_refresh\_client\_update\_subscription(...)$, where a \\null pointer returned by $strchr$ is dereferenced without a prior check. The array out-of-bounds issue was identified in the function $cmd\_show\_prompt\_history\_exec(...)$, where an index variable $type$ is used to access arrays without validating that it lies within the admissible range. Both issues can lead to undefined behavior and potential vulnerabilities within the application. 

The analysis conducted for Redis revealed multiple violations \cite{redis_issue_1, redis_issue_2}. They included array bound violated (CWE-787), invalid pointer dereference (CWE-476), null pointer dereference (CWE-476), and access to object out-of-bounds (CWE-119). Some were confirmed as false positives \cite{redis_issue_1}. Although, for the current code structure, it is indeed true, other calls to the respective function may lead to the violation found here. Again, a simple check could avoid any future problems, even if they are intentional. Moreover, if an attacker tries it directly into the source code, he should also remove such a check, which would attract the attention of the respective code maintainers. 

Anyway, a null pointer was identified in function $completionCallback(ls\rightarrow buf, \&lc)$ without confirming the non-nullity of $ls$ and $ls\rightarrow buf$ \cite{redis_issue_2}. This oversight could lead to the dereferencing of a null pointer. It does pose a risk of undefined behavior if this function is called with a null pointer. Additionally, it should be noted that null checks may be adequate for ensuring pointer validity as a pointer might seem legitimate but can be exploited to reference an invalid memory address. This nuance highlights the importance of comprehensive pointer validation prior to usage. The developers dismissed this potential issue by claiming that a function or method will never be called in a certain way that triggers a problem, despite confirmation of the issue. Consequently, this is clearly a bad practice.

Some property violations were reported to the maintainers of SQLite \cite{SQLite_issue_1, SQLite_issue_2}. These issues are related to division by zero (CWE-369), array out-of-bound (CWE-787), same object violation (CWE-469), and null pointer dereference (CWE-476). In the context of software verification, the report from LSVerifier indicates a potential vulnerability in the internal function with signature $vdbePmaWriterInit(..., int~nBuf, ...)$, in the file $vdbesort.c$, where a division by zero could occur if $nBuf$ is zero. The latter is used in a modulo operation, which, when its value is zero, results in an undefined behavior in C, leading to crashes or other unexpected behaviors. While developers might easily assert that the function $vdbePmaWriterInit()$ is never invoked with $nBuf \leq 0$, it's imperative, from a rigorous software engineering standpoint, to account for and mitigate such edge cases. This not only ensures code robustness but also preemptively addresses potential vulnerabilities. 

It is worth mentioning that, when we informed the respective developers that these results came from an automated analyzer, they started demonstrating disbelief. Indeed, the SQLite project's response to this violation reveals another prevalent behavior in the software development community. Developers often dismiss the results from static analyzers, labeling them as false positives. This perspective stems from the understanding that static analyzers frequently produce inaccurate results, leading to unnecessary alarms. The SQLite team's stance is clear: without concrete evidence in the form of an SQL script or specific code that can reproduce an issue, such reports will be disregarded. This approach, while pragmatic, is risky. Relying solely on tangible evidence might overlook potential vulnerabilities that haven't manifested yet but could be exploited in the future. An over-reliance on the historical performance of a codebase, as mentioned by the SQLite team regarding their source tree's ability to confuse static analyzers, can lead to complacency. 

It is essential to recognize that while static analyzers might produce false positives, they can also pinpoint genuine issues that might be overlooked during manual code reviews. It is crucial for developers to strike a balance. While it's unreasonable to expect teams to act on every single report from a static analyzer, completely ignoring them is not the solution either. A more collaborative approach, where the reporter and the development team work together to validate and address potential issues, can lead to more secure and robust software. After all, the ultimate goal should be to ensure the software's integrity and safeguard it from potential threats, regardless of their origin.

In summary, as seen in counterexample logs and bug-report validations, third-party libraries seem to be the biggest problem. Functions from other libraries that are called in software modules should always be carefully checked by developers before their use as they can be dangerous. Moreover, given that C programs often use pointers to access arrays and those are usually passed as arguments to functions, such a condition can bring serious security issues.

Moreover, at this point, we have enough information to tackle our research goals again. Specifically, we are able to revisit \textbf{EG2}.

\begin{tcolorbox}
As the respective software developers confirmed the problems found by LSVerifier, we can now completely answer \textbf{EG2}, confirming its feasibility for practical use. 
\end{tcolorbox}

The obtained results show that LSVerifier is well-positioned in the area of formal verification via BMC to check software vulnerabilities in large C-based software systems. This perception is also corroborated by the answers to our two research goals. Also, the prioritization algorithm enhances code analysis by function type, streamlining the identification of critical issues for efficient resolution. This verification tool is crucial because, despite developers' assurances that certain conditions are unlikely or impossible within the normal execution flow, our findings suggest otherwise. Confirmed vulnerabilities indicate that under certain manipulations, such as parameter tampering or code modification, an attacker could feasibly trigger these improbable scenarios. Such an eventuality, previously dismissed by developers, could lead to significant and severe consequences if exploited. This reinforces the importance of our methodical approach, highlighting the need for rigorous security practices even in seemingly unlikely situations.

Indeed, LSVerifier was able to successfully verify extensive software systems within a reasonable time while avoiding high memory consumption. Nevertheless, the analysis performed here shows that further work is needed to reinforce and double-check vulnerabilities so one can have an upfront confirmation of what is presented. One alternative is to develop counterexample validators useful for developers of open-source applications. Such elements can help mitigate bad practices, by showing that the refuted problems are not false positives. Without such validators, problem validation will still depend solely on extensive analysis to confirm their existence. 

Anyway, the LSVerifier's results may also be used as a mind-changing tool regarding software development practices. They show a high amount of existing vulnerabilities, which are a result of both careless coding and wrong behavior and practices, thus highlighting the need for diligent actions. 

\subsection{Threats to the Validity of Experiments}
\label{sec:threats}

We have split our analysis of threats to validity into three categories. This approach helps clarify different aspects by specifically focusing on each one at a time as follows.

\subsubsection{Benchmark selection}

We report an assessment of our vulnerability verification methodology using a set of open-source C software project benchmarks, which can be used to evaluate its effectiveness and efficiency. However, this dataset is limited within this paper's scope and its results may not be generalized to other benchmarks.

It is important to carefully select a representative set of benchmarks or dataset that can provide insights into the strengths and weaknesses of a specific scenario. Besides, we should also acknowledge the limitations of the obtained results and the potential for variation.

\subsubsection{Performance and correctness} 

The implemented strategy relies on the idea that each program function can be evaluated and such a statement might lead to accurate verification. The correctness of the verification results produced by the proposed approach may be compromised only when the assumptions made during the verification process do not accurately reflect the behavior of the program under certain conditions, e.g., inexistence of complex parallel or concurrent programming constructs. Furthermore, its performance may be impacted when dealing with benchmarks that are affected by the mentioned aspects. This is so because the verification process may need to consider all possible interleavings of the program functions, which can be computationally expensive and time-consuming. 

\subsubsection{Counterexample validation}

It is crucial to validate the counterexamples produced by LSVerifier since they serve as the foundation for checking the correctness of a target program. However, counterexample validation can be arduous, as shown here, especially when working with complex software programs or large projects. Consequently, we have conducted additional rigorous testing and analysis to independently verify the correctness of the counterexamples provided here, ensuring their precision and validity.

It is also important to perform a broader evaluation to include other scenarios and error conditions, which may reveal unexpected conditions of our tool. 

\section{Related Work}    
\label{sec:related-work}

The C programming language is widely used to develop critical software, such as operating systems, device drivers, and encryption libraries. However, it lacks protection mechanisms, leaving developers in charge of the memory and resource management procedures. Moreover, any lapse in this regard can result in odd behavior, which exposes a program to security vulnerabilities. 

Consequently, several studies have addressed this problem through the use of automatic tools for checking safety properties in C programs~\mbox{\cite{related-1, balan, related-2, Richardson2020, Rocha2020}}. Nonetheless, not all safety violations can be covered efficiently using such tools due to verification complexity and applicability. As a response, the development community began providing solutions for program testing with publicly available frameworks for static techniques, symbolic execution \mbox{\cite{baldoni2018survey}}, dynamic approaches using fuzzing and sanitization \mbox{\cite{dinesh2020retrowrite}}, abstract interpretation \mbox{\cite{rival2020introduction}}, and BMC \mbox{\cite{Clarke2004, kroening2014cbmc, 1st}}, for instance. Thereby, software developers now have many multifaceted solutions for program testing using static techniques, which allow them to customize key parameters specific to the desired test scenarios \mbox{\cite{xin2011program, situ2018vanguard, fioraldi2020fuzzing, yamaguchi2013chucky, gui2021uafsan}}.

Vorobyov, Kosmatov, and Signoles~\cite{related-2} aim to assess state-of-the-art techniques based on the performance of three runtime verification tools for C programs. For this purpose, they used approximately $700$ test cases representing security-related vulnerabilities, which were previously classified regarding memory safety. This work focused on checking C programs through dynamic analysis, a process that typically relies on code instrumentation, program execution, and the examination of one behavior at a time. The respective results indicate that dynamic analysis tools provide adequate support for detecting problems. In contrast, tools like LSVerifier perform static code analysis, which does not depend on executions and can verify multiple properties in a single run. This approach proves to be more effective when evaluating critical projects.

Fuzzing is another technique to verifying security vulnerabilities in software~\cite{Bohme2017}. In this context, B\"ohme {\it et al.}~\cite{Rocha2020} introduced Map2Check, a software verification tool that uses fuzzing, symbolic execution, and inductive invariants to assess safety properties in C programs. Furthermore, it employs the infrastructure of the low-level virtual machine (LLVM) compiler to instrument source code and monitor data during program execution. It also employs iterative deepening based on fuzzing and symbolic execution engines to verify these properties \cite{Rocha2020}. Although their experimental results demonstrate that Map2Check can be helpful when verifying pointer safety-related properties, it also exhibited low performance in the {\it MemSafety} and {\it NoOverflow} categories of SV-COMP. In addition, its evaluation is currently limited to SV-COMP's benchmarks. In contrast, our work presents results for a wide range of large open-source practical software systems.

Nie, Jiang, and Ma \cite{9251140} introduced an efficient computation tree logic (CTL) symbolic model-checking algorithm based on fuzzy logic, which addresses the state space explosion problem. Unlike conventional algorithms based on binary decision diagrams (BDDs), the proposed algorithm uses fuzzy logic to reduce the complexity of BDD computations by representing CTL formulas as fuzzy sets. It not only enhances scalability and efficiency but also inherently supports behaviors with probabilistic and temporal constraints. However, it also demonstrated limitations in representing counterexamples. Considering LSVerifier, we overcame this limitation by employing a model checker that generates counterexamples when properties are satisfied.

Alshmrany {\it et al.} \cite{alshmrany2021fusebmc} introduced FuSeBMC, a method that combines fuzzing and BMC to find security vulnerabilities in C programs. This tool is based on ESBMC, providing an efficient test generation framework. They have demonstrated the effectiveness of their approach by using it to detect SQL injection bugs in a sample web application implemented in the C language. Given that this tool introduces a fuzzing component for code analysis, this condition may result in longer verification times and extra configuration effort when compared to LSVerifier, which employs only BMC and still yields satisfactory results. In large projects, the waiting time for results can be a critical factor.

Aljaafari {\it et al.} \cite{9955513} introduced Ensembles of Bounded Model Checking with
Fuzzing (EBF). It is a method that combines BMC and Gray-Box Fuzzing (GBF) to discover software vulnerabilities in concurrent programs. The resulting tool was capable of producing up to $14.9$\% more correct verification witnesses compared to using BMC tools alone. Furthermore, this tool successfully detected a data race bug in the open-source project wolfMqtt. It was run over the benchmarks used in SV-COMP 2022. However, regarding practical software, its evaluation was limited to wolfMqtt and three other programs. In contrast, LSverifier was evaluated against various open-source large projects. Anyway, considering the performance gain achieved by adding fuzzing, it may be interesting to include this technique in future LSVerifier versions as there is potential to discover more bugs in open-source projects.

Gerking {\it et al.}~\mbox{\cite{gerking2018model}} proposed an approach to model-check the information flow security of cyber-physical systems, when represented in the form of timed automata, and used the UPPAAL model checker as underlying timed automata verification engine. This work covers aspects related to real-time behavior and asynchronous communication. However, it does not handle other security aspects addressed by LSVerifier and was not applied to large open-source software systems.

We can also mention the low-level bounded model checker (LLBMC) \mbox{\cite{merz2012llbmc}} and DIVINE \mbox{\cite{baranova2017model}}, which use BMC techniques to verify memory safety properties. LLBMC is an interesting bounded model checker based on SMT solvers; however, it has limitations related to bounded analysis and scalability, which avoid its use for large systems. DIVINE, which is an explicit-state model checker, is an efficient and versatile tool for the analysis of real-world C and C++ programs. Moreover, it provides a modular platform for the verification of real-world programs. However, a recent study \mbox{\cite{monteiro2022model}} performed an extensive evaluation and found that DIVINE needs improvements regarding performance and reliability. LSVerifier does not present such limitations. It was already applied to large software systems, as shown here, and its reliability finds support in the ESBMC's results obtained in many SV-COMP editions.

Choi\mbox{\cite{choi2011safety}} reported his experience evaluating the Trampoline operating system with the SPIN model checker. Trampoline is an open-source operating system developed for automotive devices and based on the {\it Offene Systeme und deren Schnittstellen für die Elektronik in Kraftfahrzeugen} (OSEK)/{\it Vehicle Distributed eXecutive} (VDX) international standard. Using an incremental verification approach, the authors converted the Trampoline kernel code into formal models and conducted experiments, which led to discovering a safety bug in Trampoline. This study was the first successful research case that used model-checking techniques to verify vulnerabilities in open-source software, indicating a trend in using model-checking tools in large-scale projects. LSVerifier, in turn, directly checks source code for a broad set of possible vulnerabilities without explicit manual conversion. In addition, Trampoline has only $4530$ code lines, and no scalability assessment was performed. In that sense, LSVerifier was applied to projects with millions of code lines.

To analyze vulnerabilities in heap implementations, Moritz Eckert {\it et al.} \cite{eckert2018heaphopper} proposed a tool called HEAPHOPPER. This is a novel and fully automated tool that uses model checking and symbolic execution to analyze the exploitability of heap implementations in open-source code. HEAPHOPPER has demonstrated success in checking memory allocation vulnerabilities in the presence of memory corruption. However, this work is limited to checking only memory allocation issues, which does not happen with LSVerifier, and can not handle large software systems, which is the focus of our work. The authors also mentioned the need to enhance HEAPHOPPER's performance as the number of paths to be analyzed inevitably grows.

Another approach to verify software vulnerabilities is the code browser technique. Cobra~\cite{holzmann2017cobra} uses a lexical analyzer to scan source code and create an uncomplicated linked list of lexical tokens. Structural code analysis and pattern identification can be assisted by this tool. Patterns for true positive and false positive cases are carefully defined for every syntax rule or recommendation. If the code context matches a true positive pattern, the warning is considered a true positive. If the code context matches a false positive pattern, the checked warning is deemed a false positive. In the absence of a pattern matching in the code context, it will be classified as unknown. The study performed by Thu-Trang {\it et al.}~\cite{nguyen2019multiple} showed that Cobra can identify both true positives and false positives for rules and recommendations about program syntax. LSVerifier has a more efficient methodology to check code property violations compared to this method by producing counterexamples as a standard for determining whether property violations are false positives or true positives.

Recently, Cook {\it et al.}~\cite{Cook2020} presented the use of model checkers to triage the severity of security bugs in the cloud service provider at Amazon Web Services (AWS). The authors tackled the severity of bugs discovered/reported in the Xen hypervisor \cite{xenp}, an open-source hypervisor used
in industry. In this case study, when a bug is reported, engineers should evaluate its potential threat and how quickly it needs to be fixed w.r.t. its severity. To do so, the authors have applied transformations to the original source code and implemented modifications to the C Bounded Model Checker (CBMC)~\cite{Clarke2004}, aiming to slice the program under verification and generate a reduced version of it. As a result, this model checker can easily verify the resulting program, while the obtained counterexamples can help engineers write security tests to analyze bugs further. However, several abstractions performed in the verification approach might cause the model checker to miss traces and not automatically falsify spurious traces. In this regard, it may be worthwhile to consider the use of techniques to simplify the programs analyzed by LSVerifier, making them easier to verify. Moreover, it is worth mentioning that vulnerability severity is inherently considered in the LSVerifier's prioritization strategy, which can also be modified if new severity classes or a different ranking logic must be included.

Although researchers have conducted several studies in recent years, to improve the model-checking performance related to memory safety properties, most state-of-the-art model-checking tools focus their analysis on small to medium-sized programs. Another important factor is that model-checking tools generally do not provide user-friendly outputs, and many of them produce false positives, which can become even more critical when functions are analyzed redundantly. In this work, we structured log files and spreadsheets to help organize the analysis results. Additionally, we applied a prioritization algorithm that assists users in orderly visualizing the most critical issues. 

Indeed, our study aims to close the gaps related to the verification of large software systems by creating a methodology focused on them. Moreover, we intend to bring research and practice closer by applying a tool based on this same methodology to real open-source software. Furthermore, to our knowledge, LSVerifier is the first tool capable of generating reports based on BMC counterexamples that can help software developers find issues in open-source projects.

\section{Conclusions and Future Work}
\label{sec:conclusion}

This paper presented a novel methodology that can tackle the verification of large software systems. In its context, we have also implemented this same methodology, named LSVerifier, which was used to check security bugs in real open-source C programs. 

LSVerifier scans a project's directories, individually analyzing implemented functions in C files, and flags errors related to improper pointer usage, invalid memory block access, overflows, and arithmetic errors in software projects. The associated analysis involves a model checker capable of converting C code and the specified properties into a set of boolean formulas, which are applied to a solver. The latter then provides a counterexample in case the properties are violated.

Furthermore, the LSVerifier simplifies bug identification by generating a set of report files for each execution, summarizing all software weaknesses it found. This allows users to view both the violated properties and logs generated during the respective analysis, as performed in this study by human-in-the-loop verification methodology. Additionally, we have designed and implemented a prioritization algorithm to organize the analysis so that functions with a higher potential for containing bugs or causing crashes are given priority. Consequently, this approach results in prioritized reports as critical functions are analyzed first. This is a significant contribution, given that most software model-checking tools do not address such aspects, focusing on single files or directories. In addition, this prioritization approach can favor bug-fixing procedures, which will first focus on the most important problems in a given software project. LSVerifier is freely available as open-source software under the Apache License 2.0.

We evaluated our tool and its associated algorithms using a dataset of ten large practical open-source C projects. In terms of software evaluation, we were able to find issues in databases with different sizes and target applications, ranging from tens of thousands to millions of code lines, which were confirmed by their respective maintainers (see Table \mbox{\ref{properties_analysis}}). Such achievements confirmed our initial research goals and provided evidence of the efficacy of our methodology. While its design paves the way to possibilities such as whole-system exploitation, LSVerifier is mature enough to handle large and complex open-source software, such as Wireshark, VLC, and CMake. The results obtained here demonstrate the feasibility of our approach and its potential value to the open-source software community. 

In future work, we aim to improve our verification method by supporting automated parameter selection for model checkers and employing machine learning techniques for output analysis (e.g., Large Language Models (LLMs)). Furthermore, functionalities related to interrupting and resuming partial executions might be considered given that some checks may take a long time to finish. Additionally, cluster resources will be exploited to speed up the associated verification processes. Finally, we intend to develop a new automatic validator for large software systems, allowing us to confirm the obtained counterexamples against complex projects.

\begin{acknowledgements}

The authors are grateful for the support offered by the SIDIA R\&D Institute in the Model project. This work was partially supported by Samsung, using resources of Informatics Law for Western Amazon (Federal Law No. 8.387/1991). Therefore, the present work disclosure is in accordance with as foreseen in article No. 39 of number decree 10.521/2020. This work is funded by the EPSRC grants EP/T026995/1, EP/V000497/1, EU H2020 ELEGANT 957286, Soteria project awarded by the UK Research and Innovation for the Digital Security by Design (DSbD) Programme, and Cal-Comp Electronic by the R\&D project of the Cal-Comp Institute of Technology and Innovation.

\end{acknowledgements}

%

\Urlmuskip=0mu plus 2mu
\bibliographystyle{spmpsci}      
\bibliography{citations}   


\end{document}